\documentclass[aps,prl,twocolumn,superscriptaddress]{revtex4-2}
\usepackage{amsmath}

\usepackage{booktabs}
\usepackage{float}
\usepackage{amsmath}
\usepackage{amsfonts}
\usepackage{graphicx}
\usepackage{siunitx}
\usepackage{pgffor}
\usepackage{pdfpages}

\makeatletter
\AtBeginDocument{\let\LS@rot\@undefined}
\makeatother

\begin{document}



\title{Experimental Quantification of Nonlinear Mode Coupling in Nanomechanical Resonators using Multi-tone Excitation}



\author{Chris F. D. Wattjes\textsuperscript{*,1},
        Zichao Li\textsuperscript{1},
        Minxing Xu\textsuperscript{1,2},
        Richard A. Norte\textsuperscript{1,2},
        Peter G. Steeneken\textsuperscript{1,2},
        and Farbod Alijani\textsuperscript{*,}
        }
 
\affiliation{
 Department of Precision and Microsystems Engineering, Delft University of Technology, Mekelweg 2, 2628 CD Delft, The Netherlands\\
 {\rm {\textsuperscript{2}}}Kavli Institute of Nanoscience, Delft University of Technology, Lorentzweg 1, 2628 CJ Delft, The Netherlands
}


\date{\today}

\begin{abstract}
Nonlinear modal interactions in resonant systems govern a wide range of phenomena, with broad relevance across modern physics and engineering. Yet, experimentally determining the strength of  nonlinear coupling in multimode resonators remains highly challenging.
Here, we introduce a multi-tone spectroscopy method for identifying nonlinear coupling coefficients directly from experimental data. 
Our approach employs dual-tone excitation near selected resonances which, in combination with additional probing tones at higher-order modes, generates sideband responses associated with specific modal couplings. These spectral signatures are analyzed using an inverse reconstruction procedure to quantitatively determine the corresponding nonlinear coupling strengths in the frequency domain. Using this method, we determine ten pairwise nonlinear coupling parameters across five modes of highly tensioned nanostrings, enabling the reconstruction of fully experimental, device-specific nonlinear reduced-order models. Our experimentally derived models show excellent agreement with values obtained numerically using finite element based nonlinear reduced-order models. Our method is generic and can be used for the characterization of diverse modal and intermodal couplings in mechanical and hybrid resonant systems.

\end{abstract}


\maketitle


\section{\label{sec:introduction}Introduction}
Coupling between different degrees of freedom plays an important role in shaping the dynamics of mesoscopic systems. In micro- and nanomechanical resonators, such interactions enable energy exchange \cite{bachtold2022mesoscopic} and give rise to diverse phenomena, including synchronization \cite{houri2022kuramoto}, internal resonance \cite{shoshani2021resonant, wang2022persistent, yan2022energy}, amplitude-dependent dissipation \cite{kecskekler2021tuning}, and exotic dynamical states \cite{matheny2019exotic}. Beyond their roles in fundamental studies, they also underpin a range of practical applications, such as enhancing frequency stability in resonant sensors \cite{shoshani2024extraordinary}, mechanical frequency comb generation \cite{Kekekler2022,czaplewski2018bifurcation}, coherent noise cancellation \cite{de2022coherent}, and nanomechanical computation \cite{jin2025nanomechanical,grollier2020neuromorphic}.

The estimation of nonlinear coupling parameters in resonant systems is commonly carried out by using reduced-order models (ROMs), which represent the system's motion in terms of a limited set of interacting degrees of freedom \cite{younis2011mems}. Such models are typically derived analytically~\cite{asadi2018nonlinear,das2026nonlinearmodecouplingsilicon} or numerically from Finite-Element (FE) simulations \cite{kecskekler2023multimode}. While these approaches provide valuable insight, they rely on good knowledge of material properties, boundary conditions, and geometric parameters, which can be difficult to discern experimentally, particularly at small length scales where fabrication-induced uncertainties can become significant.

To circumvent these limitations, efforts have been directed toward methods that aim to reconstruct ROMs directly from experimental data. To date, this has been commonly achieved by fitting measured system responses to analytical models, either in the frequency domain through steady-state sweeps \cite{kecskekler2021tuning,yang2015experimental,Li2025} or in the time domain via ring-down measurements \cite{chen2017direct,de2023beating, wang2022persistent}. Beyond direct response fitting, coupling terms have also been estimated by analyzing normal-mode splitting or avoided crossings between coupled modes \cite{Barakat2024}, or by using Ramsey-type spectroscopic protocols \cite{le2026precise}. However, these methods are limited to linear coupling parameters or internally resonant interactions between pairs of modes and do not readily scale to systems with multiple nonlinear interactions. As a result, an experimental framework capable of generic reconstruction of selected nonlinear coupling parameters from measurements alone remains elusive.

\begin{figure*}
    \includegraphics[width=\linewidth]{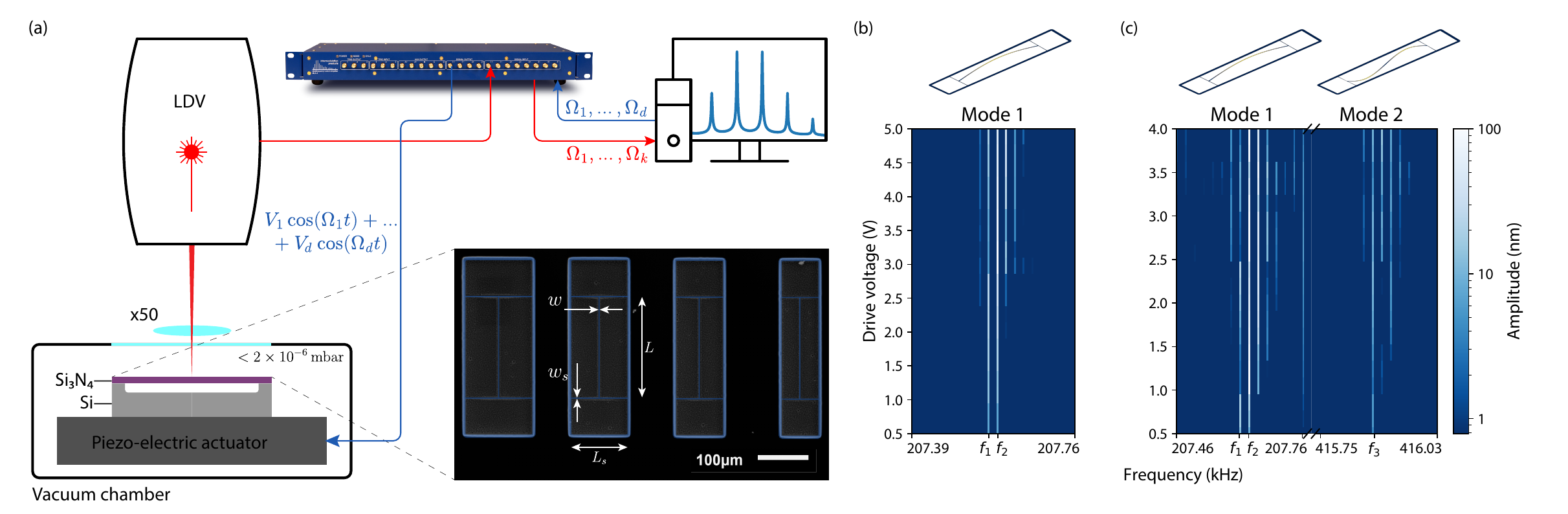}
    \caption{(a) Schematic of measurement setup, consisting of a MSA-500 Polytec laser Doppler vibrometer (LDV), Intermodulation Products Multi-frequency Lock-in Amplifier (MLA) and piezo-electric actuator to read out the motion and drive the samples at up to 32 frequencies. (b) Typical two-tone response at increasing drive voltage. Driving frequencies $f_{d=1}$ and $f_{d=2}$ are placed around the resonance frequency of the first mode (\SI{207.7}{\kilo \hertz}) with a spacing of \SI{20}{\hertz}. (c) Typical three-tone response at increasing drive voltage. Drive frequencies $f_{d=1}$ and $f_{d=2}$ are the same as in (b) while $f_{d=3}$ is added on the resonance frequency of the second mode (\SI{415.9}{kHz}).}
    \label{fig:setup}
\end{figure*}

Here, we recover nonlinear couplings in multimode resonant systems from measured sideband responses. By applying combinations of driving tones that selectively excite specific modal interactions, the method isolates individual coupling contributions, and quantifies their strength sequentially via an inverse reconstruction approach in frequency domain. Using this method, we identify ten pairwise nonlinear coupling terms in the first five vibrational modes of a nanomechanical string resonator driven in the nonlinear regime. While demonstrated here for cubic modal couplings, the method is easily extendable to other forms of dynamical interactions and to couplings between different physical degrees of freedom, providing a scalable pathway toward accurate characterization of complex nonlinearly coupled resonant systems.

\section{Experiments}

A schematic of the experimental setup is shown in Fig.~\ref{fig:setup}a. Measurements are performed on softly clamped, prestressed Si\textsubscript{3}N\textsubscript{4} nanostrings (see the inset), which we have previously shown to exhibit high $Q$-factors, Duffing-type nonlinearity, and nonlinear modal coupling~\cite{Li2024, Li2025}. These devices have a length $L$ of \SI{200}{\micro\meter}, a width $w$ of \SI{2}{\micro\meter}, support beam width $w_s$ of \SI{1}{\micro\meter} and thickness of \SI{90}{\nano\meter}. Different devices with varying support length $L_s$ are available on the same chip. To actuate the nanostrings, a piezo-electric actuator is used in combination with a multi-frequency lock-in amplifier (MLA), which generates the multi-tone driving signals. The out-of-plane vibrational response is then detected using a Polytec MSA400 laser Doppler vibrometer (LDV), and sent to the MLA which measures the phase and amplitude of the resulting sideband frequencies, allowing characterization of up to 32 frequency components simultaneously. The measurement laser is focused on the central string, at a position $L/10$ away from the support to ensure it is distant from nodal points from the first vibrational modes. To convert the measured displacement $x_n$ to modal displacement $q_n$ of mode $n$, the laser position combined with the mode shape is taken into account by using the equation $q_n=c_nx_n$, with conversion factor $c_n = 1/\sin(n \pi/10)$, assuming only one mode is active. All measurements are performed at room temperature in a vacuum chamber with a pressure below $\SI{2e-6}{\milli\bar}$ to minimize the air damping.

By sweeping the drive frequency in the neighbourhood of the natural frequencies, the resonance frequencies can be obtained. For a sample with a support length $L_s=\SI{50}{\micro\meter}$, these are \SI{207.7}{\kilo\hertz} and \SI{415.9}{\kilo\hertz} respectively for the first two out-of-plane modes. When placing two driving tones with a spacing of \SI{20}{\hertz} around the first resonance frequency, as seen in Fig.~\ref{fig:setup}b, sidebands are generated at equally spaced frequencies around the driving tones in the nonlinear regime~\cite{platz2008intermodulation}. These additional response peaks are a direct result of the Duffing-type cubic nonlinearity of the nanostring. The evolution of these sidebands as a function of drive amplitude is discussed in Supplementary Note~I. 

Interestingly, when an additional driving tone is applied at the resonance frequency of the second mode, Fig.~\ref{fig:setup}c shows that sidebands appear not only around the first mode but also around the second resonance. Since only a single tone is present near the second resonance, local nonlinear mixing cannot generate sidebands there, as at least two tones would be required. The appearance of sidebands around the second mode therefore originates from nonlinear coupling between the driven modes, providing direct spectral evidence of intermodal interaction. Supplementary Note~II further discusses the mixing caused by nonlinear coupling terms.

\section{Nonlinear Parameter Estimation}

\begin{figure*}
    \includegraphics[width=\linewidth]{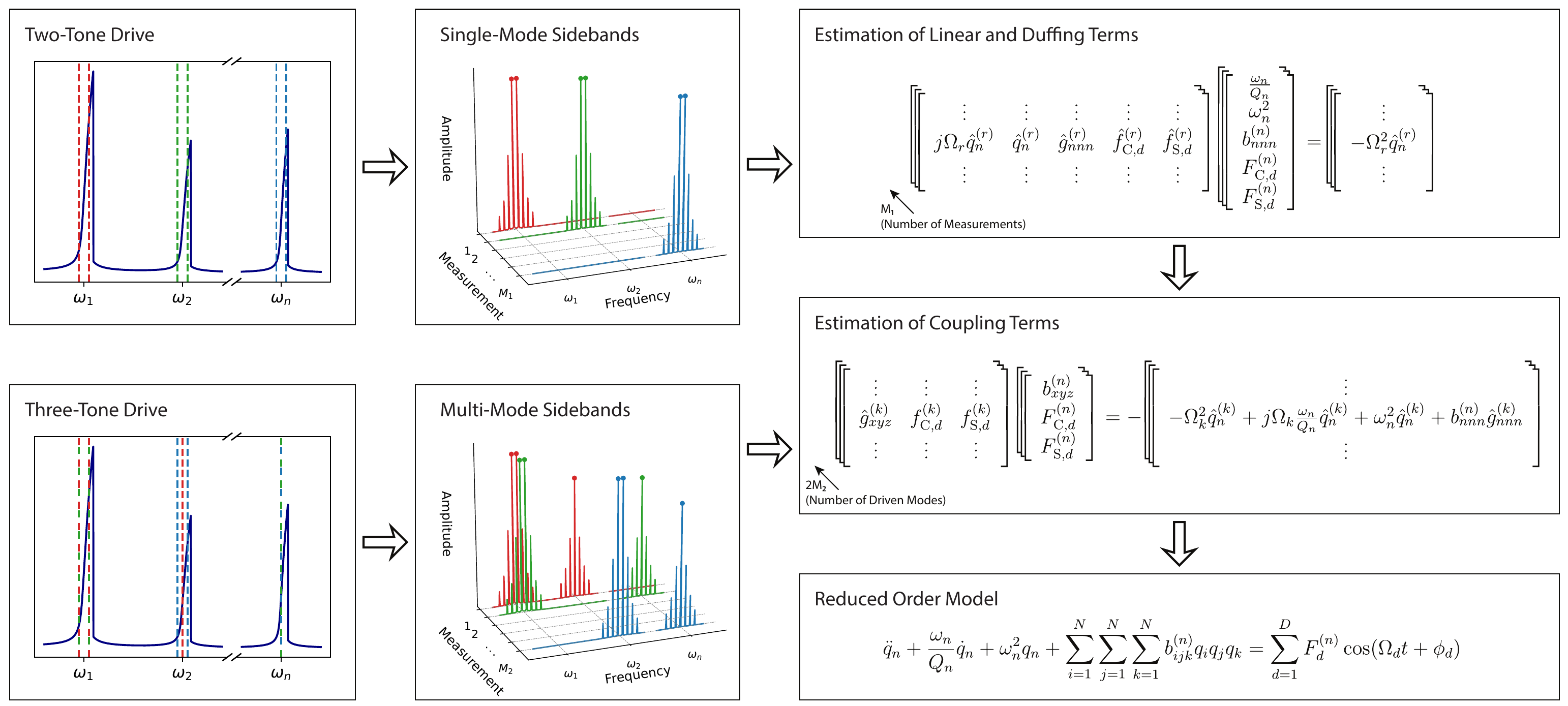}
    \caption{Schematic flow diagram of the nonlinear parameter estimation methodology. Different combinations of two-tone excitation are placed on either side of the resonance frequencies. The complex sideband amplitudes $\hat{q}^{(r)}_n$ of the single-mode sidebands are then extracted experimentally, which are used to construct Eq.~\ref{eq:one-mode-identification}, here shown in matrix form, allowing extraction of the linear parameters and Duffing nonlinearities of each mode. Subsequently,  different combinations of three-tone drive are applied, after which the generated multi-mode sidebands are measured. The obtained complex amplitides then are used to construct Eq.~\ref{eq:two-mode-identification}, again shown in matrix form, for each activated mode, enabling identification of the intermodal coupling terms. All extracted coefficients are finally combined to reconstruct a ROM describing the full nonlinear dynamics of the system. The force quadratures $F_{\rm{C},d}^{(n)}$ and $F_{\rm{S},d}^{(n)}$ are cast into the forcing term on the right hand side of the ROM. It should be noted that these terms between the two tone and three tone measurements are different.}
    \label{fig:schematic}
\end{figure*}

The key idea behind the methodology presented in this work is to use the amplitudes and phases of the generated sidebands around the resonance frequencies to determine the nonlinear coupling coefficients. For this we propose a reconstruction approach from which the strength of coupling between modes can be quantified, as shown in Fig.~\ref{fig:schematic}. After identifying the resonance frequencies by performing a low excitation frequency sweep, this approach is applied sequentially on measurements with different sets of driving tones. All single mode dynamics are analysed individually first from single mode sidebands and thereafter all coupling effects between specific combinations of modes are investigated using multi-mode sidebands. All estimated dynamical parameters are then combined to reconstruct the full ROM. 

To measure and quantify the dynamics of a single mode, two driving tones with small frequency spacing $\delta\Omega$ are applied at frequencies on either side of the resonance frequency $\omega_i$ of the mode of interest. The amplitudes $A_r$ and phases $\phi_r$ of the generated single mode sidebands are then measured at frequencies $\Omega_r = \omega_i\ +\frac{k_r}{2} \delta\Omega$, for odd integers $k_r$ in the range $\left[-R_i+1,R_i-1\right]$, such that the number of frequencies $R=R_i$ is even, and are used to obtain the complex modal amplitudes $q_i^{(r)}=c_i A_re^{\mathrm{i}\phi_r}$. 

Next, for obtaining multi-mode sidebands, an additional tone is placed directly on the higher resonance frequency $\omega_j$. Now, multi-tone sidebands appear additionally at $\Omega_r = \omega_j\ +\frac{k_r}{2} \delta\Omega$ for all even integers $k_r$ in range $\left[-R_j,R_j\right]$, where $R_j$ is odd, such that the total number of frequencies $R=R_i+R_j$. To obtain the complex modal displacements $q_n^{(r)}$ for $n=i,j$, we assume the observed sidebands to correspond to the mode with the spectrally closest resonance frequency. Therefore, the modal displacement can simply be obtained as $q_n^{(r)}=c_n A_re^{\mathrm{i}\phi_r}$ with $n$ being $i$ or $j$ depending on which resonance frequency is closest to $\Omega_r$, while the modal displacement for the other mode is assumed to be zero. 

In order to quantify all linear and nonlinear single-mode terms, first $M_1=N$ measurements with different two-tone driving combinations are conducted to activate each mode separately, where $N$ is the number of modes participating in the dynamics. To activate and quantify all two-mode couplings, every combination of modes should be activated, leading to $M_2=\frac{1}{2} N(N-1)$ measurements with different three-tone driving combinations. Together, these measurements provide the data required for identifying all parameters describing up to two-mode interactions in the equations of motion.

To obtain the nonlinear coefficients from the described measurements, we propose an inverse reconstruction approach. To that end, here we generalize the quantification to an arbitrary number of $N$ modes, and use the following nonlinear ROM:
\begin{equation}\label{eq:reduced-order-model}
\begin{split}
    &\ddot{q}_n + \frac{\omega_n}{Q_n} \dot{q}_n + \omega_n^2 q_n + \sum_{i=1}^{N} \sum_{j=1}^{N} \sum_{k=1}^{N} b_{ijk}^{(n)} q_i q_j q_k \\&\quad = \sum_{d=1}^{D} F^{(n)}_{d} \cos\!\left(\Omega_d t + \phi_d \right), \qquad n=1,...,N,
\end{split}
\end{equation}

where $q_n$ describe the modal displacements, $\omega_n$ and $Q_n$ are the linear angular resonance frequency and Q-factor of mode $n$ and the parameters $b^{(n)}_{ijk}$ represent cubic stiffness coefficients. The modal forces $F^{(n)}_{d}$ represents the multi-tone drives placed at frequencies $\Omega_d$ at, or in the spectral neighbourhood of, the resonance frequencies. Furthermore, $D$ is the number of drive tones in the experiments. Here, it is assumed that quadratic nonlinearities can be neglected due to symmetry of the nanostrings in the out-of-plane direction \cite{Li2025}.

Based on the experimentally observed sidebands at frequencies $\Omega_r$, we can write the solution of Eq.~\ref{eq:reduced-order-model} in the form $q_n(t)=\sum_{r} \hat{q}^{(r)}_n e^{\mathrm{i}\Omega_rt}$ where $\hat{q}^{(r)}_n$ are the complex amplitudes of the sidebands generated at frequencies $\Omega_r$, assuming that displacement at other frequencies or their influence on displacement at $\Omega_r$ are negligible. 
By inserting this solution into Eq.~\ref{eq:reduced-order-model} and collecting terms proportional to each sideband frequency $\Omega_r$, the equations of motion can be projected onto the sideband frequencies. This results in a set of $N \cdot R$ equations for every mode $n$ and frequency $\Omega_r$:
\begin{equation}\label{eq:equation-of-motion-per-frequency}
\begin{split}
    &\Omega_r^2\hat{q}^{(r)}_n + \frac{\omega_n}{Q_n} j\Omega_r\hat{q}^{(r)}_n + \omega_n^2 \hat{q}^{(r)}_n + \sum_{i=1}^{N} \sum_{j=1}^{N} \sum_{k=1}^{N} b_{ijk}^{(n)} \hat{g}^{(r)}_{ijk} 
    \\&\quad= \sum_{d=1}^{D} F^{(n)}_{\mathrm{C},d} \hat{f}^{(r)}_{\mathrm{C},d} + F^{(n)}_{\mathrm{S},d} \hat{f}^{(r)}_{\mathrm{S},d}, \qquad 
    \left\{
    \begin{array}{c}
        r = 1,\ldots,R, \\
        n = 1,\ldots,N,
    \end{array}
    \right.
\end{split}
\end{equation}

where $\hat{g}^{(r)}_{ijk}$ represents the complex amplitude of the nonlinear displacement product $q_i(t)q_j(t)q_k(t)$ at $\Omega_r$. This is calculated by reconstructing the time domain displacement as $q_n=\sum_r \hat q_n^{(r)} e^{\mathrm{i}\Omega_rt}$ for $n=i,j,k$, from the measured complex modal amplitudes. Then, by multiplying the displacements and performing a Fourier transform at frequency $\Omega_r$, such that $\hat{g}^{(r)}_{ijk} = \mathcal{F}^{(r)}\left\{q_iq_jq_k\right\}$.
Furthermore, the driving force $F_d^{(n)}$ is split into an in-phase and out-of-phase component, represented by $F^{(n)}_{\mathrm{C},d}$ and $F^{(n)}_{\mathrm{S},d}$, respectively. Their contributions at frequency $\Omega_r$ are defined by the drive and sideband frequencies as $\hat{f}_{\mathrm{C},d}^{(r)}=\mathcal{F}^{(r)}\{\cos(\Omega_dt)\}$ and $\hat{f}_{\mathrm{S},d}^{(r)}=\mathcal{F}^{(r)}\{\sin(\Omega_dt)\}$, which are both zero when $\Omega_d\neq\Omega_r$. The derivation of Eq~\ref{eq:equation-of-motion-per-frequency} is discussed in more detail in Supplementary Note~III.

Since the complex amplitudes, sideband frequencies and driving frequencies are known from experiments, the unknown quantities in Eq~\ref{eq:equation-of-motion-per-frequency} are the natural frequencies $\omega_n^2$, linear damping $\frac{\omega_n}{Q_n}$, driving force components $F^{(n)}_{\mathrm{C},d}$ and $F^{(n)}_{\mathrm{S},d}$, and nonlinear stiffness $b_{ijk}^{(n)}$. Once these parameters are identified, all linear coefficients, nonlinear coefficients, force amplitudes, and force phases in the equation of motion (Eq.~\ref{eq:reduced-order-model}) can be determined. Consequently, parameter estimation reduces to solving Eq.~\ref{eq:equation-of-motion-per-frequency} for the unknowns using a least-squares optimisation procedure.

In principle, all linear, Duffing, and intermodal coupling parameters could be obtained simultaneously by solving a single optimisation problem using measurements where many modes are driven at once. In practice, however, the success of such an approach depends strongly on the chosen driving frequencies and amplitudes, and can be dominated by the stronger contributions to the response, namely the linear terms and Duffing nonlinearities, which could lead to less accurate estimation of the coupling terms. 
Performing a selected set of measurements, driving specific modes with combinations of tones mitigates this, which is the reason we opt for this strategy. Adding sequential analysis of the driven response, starting with single driven modes, allows for quantification of linear terms and single-mode Duffing nonlinearities before the effects of coupling are introduced. These estimates can be considered known when moving to the quantification of coupling coefficients, improving the accuracy of the optimization.

\begin{figure*}
    \includegraphics[width=\linewidth]{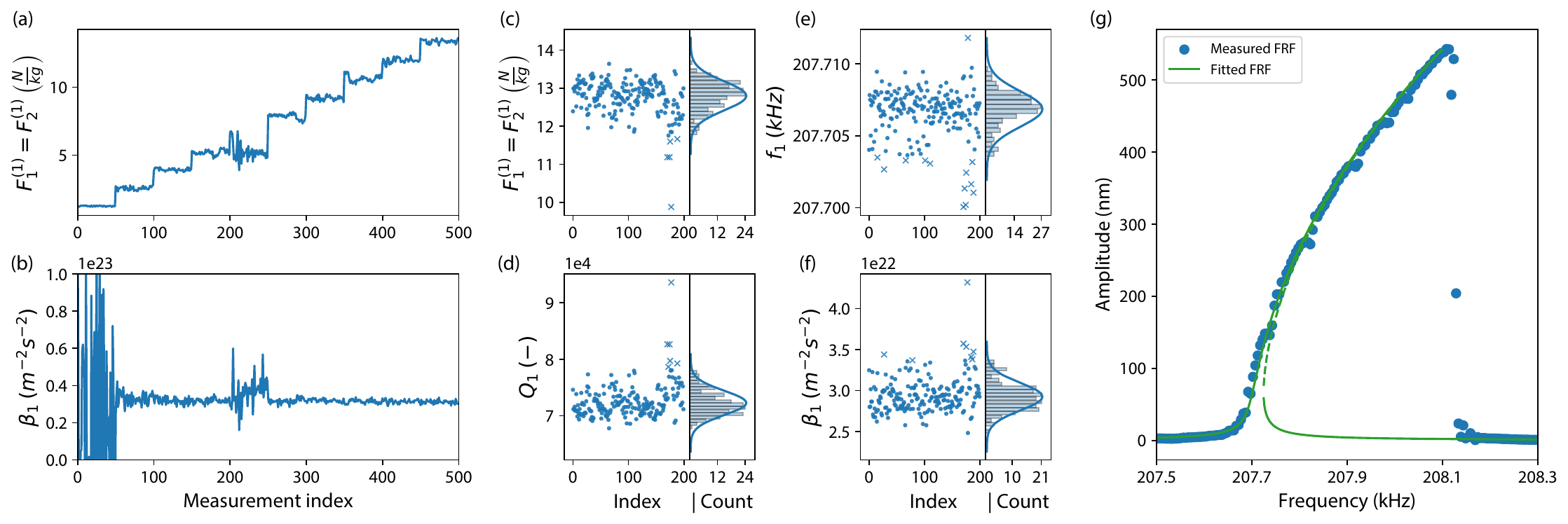}
    \caption{Nonlinear parameter estimation with multi-tone excitation. (a) The amplitude of the multi-tone excitation is increased incrementally after every 50 measurements of the sidebands, indicated by the measurement index and (b) the Duffing term $\beta$ per measurement index is quantified. (c-f) Estimated parameters for 200 consecutive measurements at the same excitation level. A normal distribution shown for a range of $\pm4\sigma$ is fitted to the histogram of the estimates, with outliers (blue x) that fall outside the interquartile range not considered. (g) Comparison between measured frequency response and reconstructed frequency response using the median of the results seen in (c-f).}
    \label{fig:duffing-results}
\end{figure*}

In the first step of the procedure, the analysis of single mode response, Eq.~\ref{eq:equation-of-motion-per-frequency} simplifies drastically. The nonlinear contribution reduces to the intrinsic cubic term $b^{(i)}_{iii}\,\hat{g}^{(i)}_{iii}$, allowing the equation to be simplified by excluding all other nonlinear terms to 
\begin{equation}\label{eq:one-mode-identification}
\begin{split}
    &\Omega_r^2\hat{q}^{(r)}_i + \frac{\omega_i}{Q_i} j\Omega_r\hat{q}^{(r)}_i + \omega_i^2 \hat{q}^{(r)}_i + b_{iii}^{(i)} \hat{g}^{(r)}_{iii} 
    \\&\quad= \sum_{d=1}^{D} F^{(i)}_{\mathrm{C},d} \hat{f}^{(r)}_{\mathrm{C},d} + F^{(i)}_{\mathrm{S},d} \hat{f}^{(r)}_{\mathrm{S},d}, \qquad 
    r = 1,\ldots,R.
\end{split}
\end{equation}

Optimising for the unknowns and repeating this procedure for all $N$ modes yields the linear modal parameters $\omega_n$, $Q_n$, the Duffing coefficients $b^{(n)}_{nnn}$, and the effective forcing terms. 

The next step in the procedure is to identify coupling terms between two modes $i$ and $j$. Here , we consider cubic couplings of the form $b^{(i)}_{ijj}\, q_i q_j^2$ and $b^{(j)}_{iij}\, q_i^2 q_j$, consistent with earlier works on nanostrings~\cite{Li2025}. Therefore, only these terms, along with the intrinsic cubic term for each mode, are considered in the nonlinear response. Eq.~\ref{eq:equation-of-motion-per-frequency} is thus simplified by excluding all other nonlinear terms. Next to this, optimising for the nonlinear coupling coefficients is simplified by using previously obtained linear terms and Duffing coefficients.

To improve the physical meaningfulness of the estimates, we consider the underlying nonlinear interaction potential which links the coupling terms in the equations of motion of modes $i$ and $j$. For these couplings, the nonlinear potential is defined as $U_{\rm{nl}}=\frac{1}{2}\gamma_{i,j}q^2_iq^2_j$ with $\gamma_{i,j}$, from which we obtain $b^{(i)}_{ijj}=b^{(j)}_{iij}=\gamma_{i,j}$ by taking the derivative with respect to each modal displacement individually. By inserting this constraint into Eq.~\ref{eq:equation-of-motion-per-frequency} and fitting for $\gamma_{i,j}$, the fitted parameters are forced to be consistent with the assumed potential. To that end, the equations that are used for identifying coupling terms are adapted to:
\begin{subequations}\label{eq:two-mode-identification}
\begin{align}
\begin{split}
    &\Omega_r^2\hat{q}^{(r)}_i + \frac{\omega_i}{Q_i} j\Omega_r\hat{q}^{(r)}_i + \omega_i^2 \hat{q}^{(r)}_i + b_{iii}^{(i)} \hat{g}^{(r)}_{iii} + \gamma_{i,j} \hat{g}^{(r)}_{ijj} 
    \\[-2pt]&\quad= \sum_{d=1}^{D} F^{(i)}_{\mathrm{C},d} \hat{f}^{(r)}_{\mathrm{C},d} + F^{(i)}_{\mathrm{S},d} \hat{f}^{(r)}_{\mathrm{S},d}, \qquad 
        r = 1,\ldots,R, \\[6pt]
\end{split}\\
\begin{split}
    &\Omega_r^2\hat{q}^{(r)}_j + \frac{\omega_j}{Q_j} j\Omega_r\hat{q}^{(r)}_j + \omega_j^2 \hat{q}^{(r)}_j + b_{jjj}^{(j)} \hat{g}^{(r)}_{jjj} + \gamma_{i,j} \hat{g}^{(r)}_{iij} 
    \\[-2pt]&\quad= \sum_{d=1}^{D} F^{(j)}_{\mathrm{C},d} \hat{f}^{(r)}_{\mathrm{C},d} + F^{(j)}_{\mathrm{S},d} \hat{f}^{(r)}_{\mathrm{S},d}, \qquad 
        r = 1,\ldots,R, \\
\end{split}
\end{align}
\end{subequations}

where the only unknowns are coupling term $\gamma_{i,j}$ and the forcing terms $F^{(n)}_{\mathrm{C},d}$ and $F^{(n)}_{\mathrm{S},d}$. These terms can again be obtained by performing a least squares optimisation procedure. By repeating this for different combinations of modes coupling strengths between those modes can be quantified and  the experimentalROM can be reconstructed. We note Eq~\ref{eq:one-mode-identification} and Eq~\ref{eq:two-mode-identification} can also be displayed in matrix form, which could be beneficial in practice, for instance when performing least squares optimisation. These matrix forms and their derivations are shown in Supplementary Note~IV. 

It is worth mentioning that the proposed framework is expandable to different types of nonlinearities, such as quadratic stiffness and nonlinear damping. Additionally, the framework naturally scales to higher-order interactions. For instance, three-mode coupling can be accessed by introducing an additional driving tone at a higher resonance, thereby activating three modes simultaneously, which in general requires $M_3=\frac{1}{6}N(N-1)(N-2)$ excitation combinations. Furthermore, coupling processes involving larger sets of modes can be investigated by extending the multi-tone excitation scheme, highlighting the scalability and generality of the approach.

\section{Results}\label{sec:results}

\begin{figure*}
    \includegraphics[width=\linewidth]{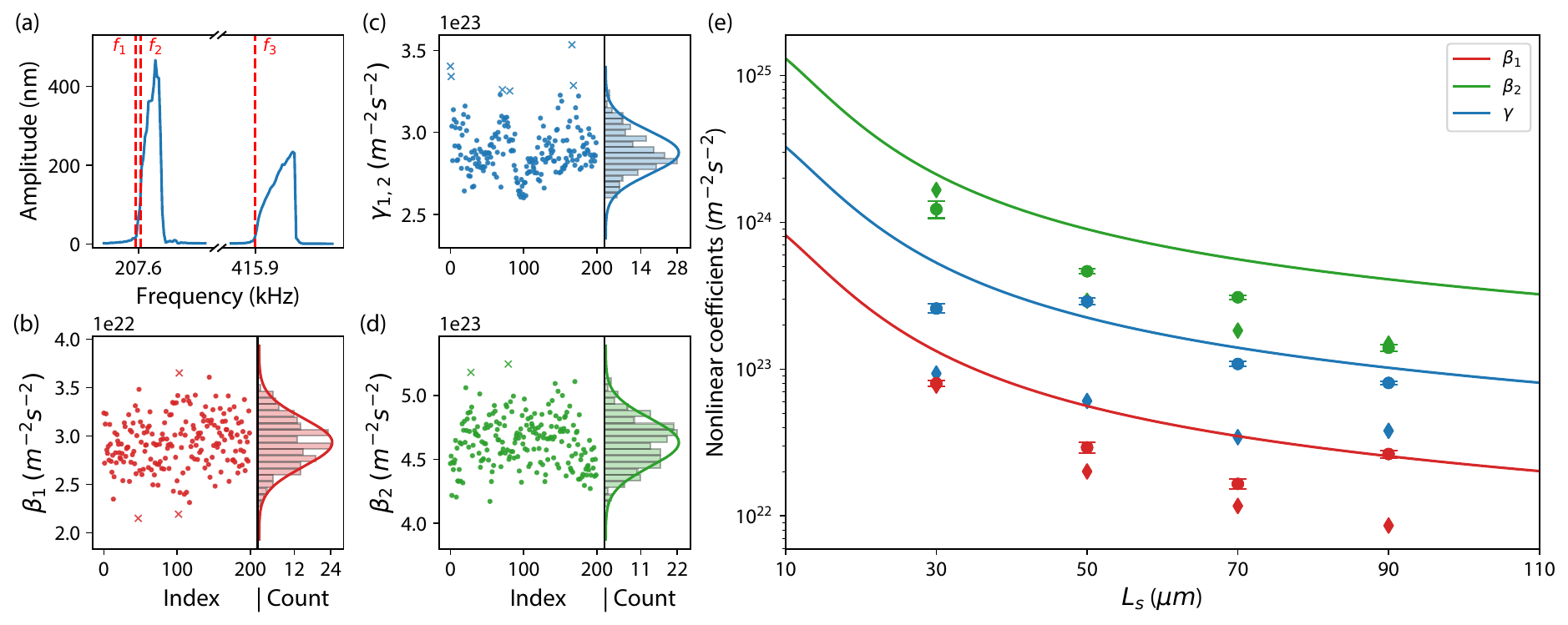}
    \caption{Nonlinear parameter estimation of two coupled degrees of freedom of the nanostring. (a) Single tone frequency sweep response for mode 1 and 2 of a sample with $L_s=\SI{50}{\micro\meter}$. Driving frequencies used for the quantification of the coupling strength are shown by the red dashed lines. (b, c, d) Estimated values for the nonlinear parameters in the equations of motion. The responses have been measured 200 times sequentially, with a normal distribution shown for $\pm4\sigma$ fitted to the histogram of the estimates. Outliers (blue x) that fall outside the interquartile range are not considered. (e) Comparison of estimated values of the nonlinear parameters in different samples obtained in this work (dots with error bars), and from theory (lines) and experiments (diamonds) shown in~\cite{Li2025}.}
    \label{fig:coupling-results}
\end{figure*}

To demonstrate the method experimentally, we apply it to Si\textsubscript{3}N\textsubscript{4} nanostrings driven with multi-tone excitation, as shown in Fig.~\ref{fig:setup}. As a first step, and to validate the approach, we focus on characterizing the Duffing nonlinearity of the fundamental mode, such that $M_1=1$ and $M_2=0$. In the following, the Duffing coefficient $b^{(n)}_{nnn}$ from Eq.~\ref{eq:one-mode-identification} is denoted by $\beta_n$ for simplicity. We note that accurate estimation of the nonlinear parameters requires driving the resonator to sufficiently large amplitudes such that it operates within the nonlinear regime. This is shown in Fig.~\ref{fig:duffing-results}a-b, where the driving voltage is increased from \SI{0.5}{\volt} to \SI{5.0}{\volt} in steps of \SI{0.5}{\volt}, thereby driving the resonator to larger amplitudes. In these measurements, the driving tones are placed symmetrically around the fundamental resonance frequency with a spacing of \SI{20}{\hertz}. The response near the driving frequencies is then measured 50 times for each driving voltage using the MLA with a bandwidth of \SI{5}{\hertz}, leading to an integration time of \SI{200}{\milli\second} per measurement index. The same tone spacing, bandwidth and methodology are used for all subsequent measurements. 

The result of the extractions of linear modal force and the Duffing parameter for the fundamental mode is shown in Fig.~\ref{fig:duffing-results}a-b, where increasing the drive amplitude leads to a reduced spread in the estimated Duffing coefficient $\beta_1$ of the fundamental mode. At low excitation amplitudes, no reliable estimate of $\beta_1$ can be obtained, indicating that the resonator has not yet entered the nonlinear regime. The increased noise observed between measurement indices 200 and 250 originates from a transient disturbance caused by the vacuum pump. Despite this perturbation, the extracted value of $\beta_1$ remains only weakly affected, demonstrating the robustness of the method. A numerical analysis of the sensitivity of the nonlinear parameter estimation to noise is further provided in Supplementary Note~V. Building on this observation, repeated measurements (200 realizations) of the $Q$-factor, resonance frequency ($f_1=\omega_1/2\pi$), and dual-tone amplitudes $F^{(1)}_1$ and $F^{(1)}_2$ are performed at a drive voltage of \SI{5.0}{\volt}, with the resulting distributions shown in Fig.~\ref{fig:duffing-results}(c--f). The narrow spread of the reconstructed parameters indicates a high level of repeatability and confirms that measurement noise introduces only limited variability in the estimated parameters.

\begin{figure*}
    \centering
    \includegraphics[width=\linewidth]{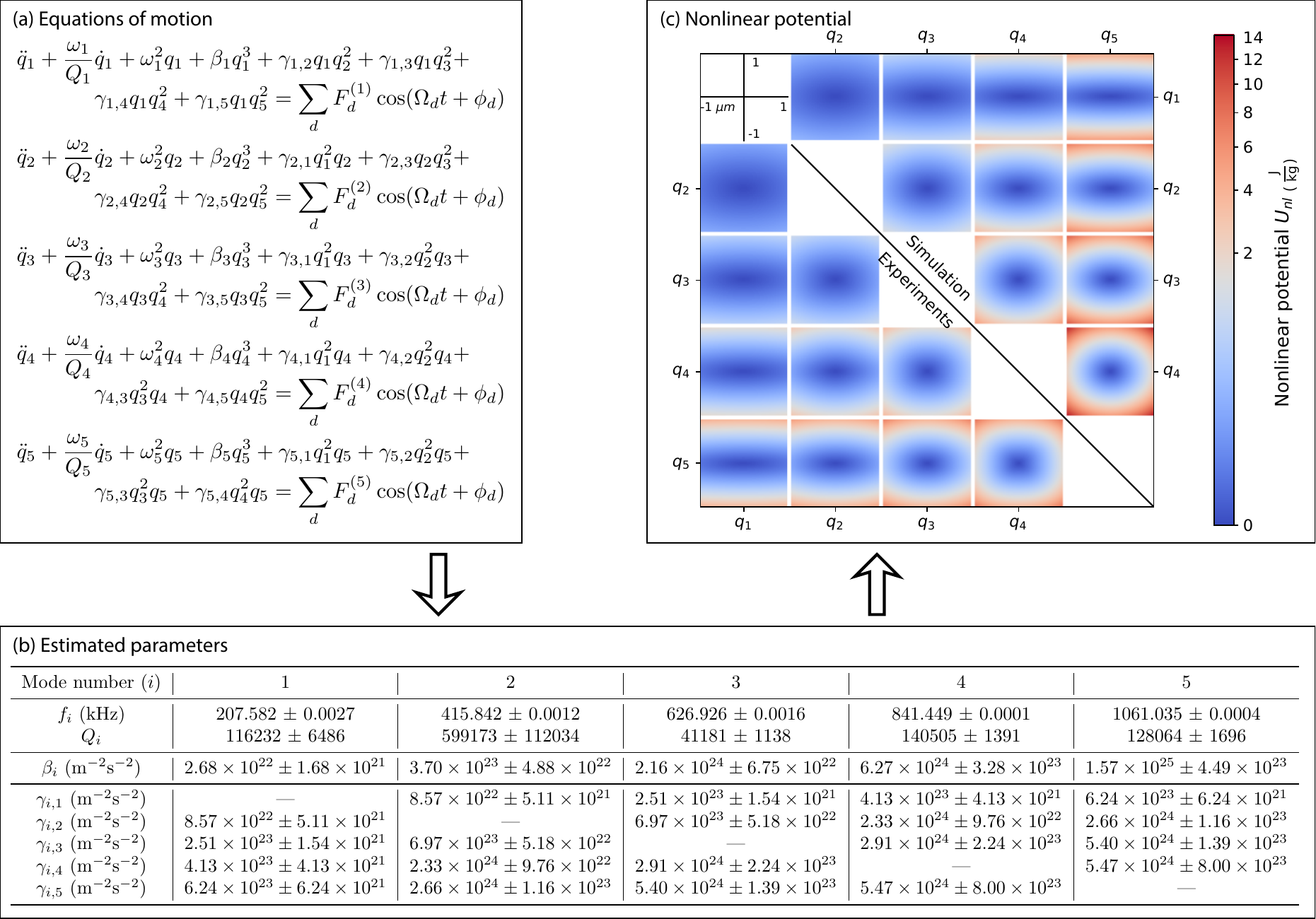}
    \caption{Nonlinear parameter estimation for the first five modes of a string with $L_s=\SI{50}{\micro\meter}$. 
    (a) Nonlinear equations of motion for the first five modes of the device, which include Duffing nonlinearity and pairwise coupling coefficients. 
    (b) All 10 linear and 15 nonlinear coefficients which are estimated by measuring the response to 15 different driving combinations, where $\gamma_{i,j}=\gamma_{j,i}$. (b) Using the estimated parameters, the nonlinear interaction potential between different modes is obtained. The experimental reconstruction (bottom left triangle) is compared to simulated results (top right triangle, values are shown in Supplementary Note VI) on a range of $\pm \SI{1}{\micro\meter}$ of modal displacement for each combination of modal coordinates.}
    \label{fig:nonlinear-potential}
\end{figure*}

To further validate the extracted Duffing parameter, a single-tone frequency sweep is also performed around the fundamental resonance at the same drive amplitude used during the multi-tone excitation. Using the median values obtained from Fig.~\ref{fig:duffing-results}(c--f), the corresponding nonlinear frequency response is reconstructed and compared with the experimentally measured Duffing response in Fig.~\ref{fig:duffing-results}(g). The excellent agreement between the reconstructed curve and the measured nonlinear resonance demonstrates that the sideband based reconstruction can accurately capture both the linear and nonlinear dynamics of a single-mode resonator.

Next, we demonstrate the reconstruction of an experimental ROM for two coupled modes of the same nanostring. Following the described procedure, we begin by quantifying the linear parameters and Duffing nonlinearity of each mode individually. Two driving tones are first placed symmetrically around the resonance of mode~1, at $f_{d=1,2}=\SI{207.7}{\kilo\hertz}\pm\SI{20}{\hertz}$ with an amplitude of \SI{4.0}{\volt}, to determine its resonance frequency, $Q$-factor, and Duffing coefficient. The same procedure is then repeated for the second mode, driven at $f_{d=1,2}=\SI{416.0}{\kilo\hertz}\pm\SI{20}{\hertz}$ at the same drive level, such that $M_1=2$.

Having established the single-mode coefficients, we proceed by applying three-tone excitation with $M_2=1$. Here, two tones are placed symmetrically around the first resonance ($f_{d=1,2}=\SI{207.7}{\kilo\hertz}\pm\SI{20}{\hertz}$) while a third tone is applied at the resonance frequency of the second mode ($f_{d=3}=\SI{416.0}{\kilo\hertz}$), as illustrated in Fig.~\ref{fig:coupling-results}a. The multi-mode sidebands that appear are used to construct Eq.~\ref{eq:two-mode-identification} and extract the coupling coefficient $\gamma_{1,2}=b^{(1)}_{122}=b^{(2)}_{112}$.

Furthermore, to assess the generality of the method, we repeat the procedure with $M_1=2$ and $M_2=1$ measurements on nanostrings with different support lengths $L_s=\SI{30}{\micro\meter}$, $\SI{70}{\micro\meter}$, and $\SI{90}{\micro\meter}$. The extracted nonlinear parameters for all devices are summarized in Fig.~\ref{fig:coupling-results}e. The obtained results are compared with theoretical predictions based on a simplified continuum mechanics model, as well as with experimental values derived from the change in slope of the Duffing backbone curve under single-tone frequency sweeps, both reported in~\cite{Li2025}. All values in Fig.~\ref{fig:coupling-results}e are shown in Supplementary Note~VI. Close agreement is observed between the coupling and Duffing parameters extracted in the present work and those obtained from both the analytical model and the backbone-slope analysis. 


Finally, to demonstrate the scalability of the approach, we reconstruct a nonlinear ROM involving the first five vibrational modes of the nanostring. After performing $M_1=5$ measurements and determining the linear parameters and Duffing nonlinearities for each mode following the procedure explained earlier, pairwise coupling coefficients are extracted under the assumption of a nonlinear interaction potential of the form $U_{\mathrm{nl}}=\sum \frac{1}{2}\gamma_{i,j}q_i^2q_j^2$~\cite{Li2025}. To identify all associated interaction terms, $M_2=10$ measurements are performed, one for every pair of modes, selectively activating them through multi-tone excitation. The complete set of extracted nonlinear parameters, including five Duffing coefficients and ten intermodal coupling terms, is summarized in Fig.~\ref{fig:nonlinear-potential}, providing a fully experimental reconstruction of a five-mode nonlinear interaction potential directly from sideband measurements. Here, Fig.~\ref{fig:nonlinear-potential}c presents the reconstructed nonlinear potential arranged as a matrix, where each row and column correspond to a specific modal coordinate. The lower triangular panel here shows the potentials obtained from the experimental reconstruction, while the upper triangular panel displays the corresponding potentials derived from FE-based reduced-order modelling of the nanostring \cite{Li2025}, for which the parameters are shown in Supplementary Note~VI. The close agreement between experiments and simulations across all modal pairs demonstrates that the experimentally extracted parameters accurately capture both the structure and magnitude of the nonlinear interactions. Overall, this comparison highlights that the proposed method provides a direct pathway toward data-driven reconstruction of high-dimensional nonlinear ROMs in complex multimode resonant systems.

\section{Discussion}

To summarize, we developed an experimental frequency-domain framework that reconstructs nonlinear ROMs directly from intermodulation spectra. Applied to Si\textsubscript{3}N\textsubscript{4} nanostrings, the method enabled identification of Duffing nonlinearities, intermodal couplings, and reconstruction of a five-mode nonlinear interaction potential in close agreement with simulations. Unlike commonly used experimental approaches to estimate coupling, the presented method accesses nonlinear interactions without requiring high drive amplitudes and carefully chosen frequency steps to achieve coupled dynamics. It also avoids reliance on higher-order autoparametric resonances~\cite{Li2025}, operation in mode-splitting~\cite{Barakat2024} or internal resonance regimes~\cite{chen2017direct}. Instead, here, coupling signatures emerge directly from nonlinear mixing between multi-tone drives, allowing specific modal interactions to be activated in a controlled and repeatable manner. An additional advantage lies in the frequency-domain formulation of the approach. In contrast to many data-driven identification techniques that rely on time-domain measurements \cite{Brunton2016,cenedese2022data,bosso2026machinelearningframeworkuncovering} and thus can be sensitive to noise or transient disturbances, the present approach extracts parameters from steady-state spectral components. Consequently, the estimated nonlinear coefficients are less affected by experimental perturbations, as illustrated in Fig.~\ref{fig:duffing-results}. Moreover, the ability to obtain ROMs directly from experiments offers a complementary pathway to analytical or FE-based reduced-order modelling of nonlinear dynamics \cite{das2026nonlinearmodecouplingsilicon, kecskekler2023multimode, jain2022compute}. While such numerical methods capture the underlying mechanics with high fidelity, their predictive accuracy can be limited by uncertainties at the micro or nanoscale. Experimental construction of nonlinear interaction parameters therefore provides a practical route toward device-specific models that can guide the design and optimization of multimode micro and nanoresonators~\cite{li2025finite, pozzi2025topology}. 

Furthermore, although the present work focused on pairwise modal interactions, a natural extension is the investigation of higher-order processes such as three-mode mixing. In the nanostrings studied here three-mode interactions appear to be small~\cite{Li2025} and as discussed in Supplementary Note~II were not observed. We however expect that by performing measurements at frequencies that do not lie around the driving frequencies, such coupling terms could, in principle, be quantified. Although, this would necessitate the incorporation of a modal decomposition procedure, as the assumption that the motion is dominated by the mode with the spectrally closest resonance frequency would no longer be valid. Applying the framework to systems where resonant three-mode processes occur~\cite{houri2020demonstration} would provide an interesting direction for future studies and further test the scalability of the approach. Finally, even though demonstrated here for cubic nonlinear couplings in nanomechanical resonators, the methodology is not restricted to this interaction class. The same frequency-domain strategy could be extended to systems exhibiting linear \cite{le2026precise} or quadratic coupling \cite{Kekekler2022} interactions, or hybrid systems such as optomechanical devices \cite{yao2025longrange}. In this broader context, the proposed approach may provide a general experimental tool for probing and quantifying complex interaction networks in resonant systems.

\section{Authors contribution}
{C.F.D.W. developed the nonlinear parameter estimation algorithm; M.X. and R.A.N. fabricated the samples; The measurments were performed by C.F.D.W. with input from Z.L.; Data analysis and interpretation were done by C.F.D.W and F.A.; The project was supervised by P.G.S and F.A.; and the manuscript was written by C.F.D.W. and F.A. with input from all authors.}

\section{Acknowledgements}

Funded by the European Union (ERC
Consolidator, NCANTO, 101125458). Views and opinions expressed are, however, those of the author(s) only and do not necessarily reflect those of the European Union or the European Research Council. Neither the European Union nor the granting authority can be held responsible for them. C.F.D.W. acknowledges the fruitful discussions with Dr. Alberto Martin-Perez, Mrs. Enise Kartal, Mr. Kushal Swamy, and Dr. Daniel Forchheimer from Intermodulation Products.

\bibliography{references}

\clearpage



\section*{Supplementary material}

# Supplementary Information: Experimental Quantification of Nonlinear Mode Coupling in Nanomechanical Resonators using Multi-tone Excitation

Chris F. D. Wattjes[*,1], Zichao Li[1], Minxing Xu[1,2], Richard A. Norte[1,2], Peter G. Steeneken[1,2], and Farbod Alijani[*,1]

[1]*Department of Precision and Microsystems Engineering,
Delft University of Technology, Mekelweg 2, 2628 CD Delft, The Netherlands*
[2]*Kavli Institute of Nanoscience, Delft University of Technology, Lorentzweg 1, 2628 CJ Delft, The Netherlands*

(Dated: April 15, 2026)

## SUPPLEMENTARY NOTE I. NONLINEAR BEHAVIOUR OF A DUAL-TONE DRIVEN DUFFING RESONATOR

When driving a Duffing oscillator with two-tone drive, sidebands are generated at equally spaced frequencies around the driven tones. Naturally, these sidebands become more apparent as the system is driven further into the nonlinear regime. Additionally, more complex behaviour of the sidebands is observed as drive is increased further, showing multiple jumps in amplitude. Fig. S1a shows the response of the sample with a support length of $L_s = 50\,\mu$m which is used in the main text. The sample is driven around its fundamental resonance at $f_{d=1,2} = 207.6$ kHz $\pm 20$ Hz, and the response is measured at the driven frequencies and the two closest sidebands. As the drive voltage is swept up, the amplitude of the two driven tones increases linearly, until a sudden jump in amplitude for the higher tone and a jump down for the lower tone. At higher drives, more jumps are visible, while the amplitude of the higher driving tone decreases for larger drive. To verify the experimentally observed dynamics, numerical simulations are performed using the parameters of Fig. 3 in the main text. The results, presented in Fig. S1b, show a close agreement with the experimental results, also showing multiple jumps in amplitude. While the dynamics of two-tone driven Duffing resonators have been investigated [1], the behaviour seen here in the frequency domain has not been reported before.

It should be noted that all measurements performed for identification in the main text use a drive voltage of 4 V or 5 V. Here, no jumps are observed for the first mode. We note that as long as the response remains periodic, the methodology presented in the main text is valid and the identification procedure can be still applied.

## SUPPLEMENTARY NOTE II. FREQUENCY MIXING FOR NONLINEAR TERMS

The measurement frequencies $\Omega_r$ are chosen by analysing the spectral mixing products generated by the nonlinear interaction terms. To show the procedure, we apply the harmonic balancing method on a number of coupled nonlinear forcing terms. From these, it becomes clear why only certain coupling terms are activated by the type of experiments performed in the main text.

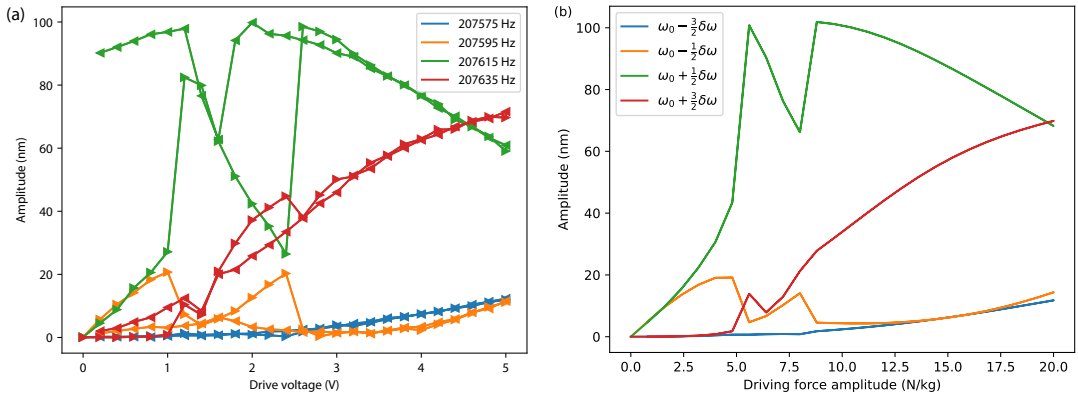

FIG. S1. Experimentally and numerically obtained response of a nanostring driven with two tones close to its fundamental resonance $\omega_0$ at different drive power. Both subfigures show the reponse at the driven tones and the two closest sidebands, colors match between subfigures. In (a), right facing triangles (>) and left facing traingles (<) indicate the path when drive power is swept up and down respectively.

Starting with single mode Duffing nonlinearity, we assume the motion to be $q_i(t) = \sum_d X_d \cos(\omega_d t)$, which in the case of single mode identification would be $q_1(t) = X_1 \cos((\omega_1 + \delta\omega)t) + X_2 \cos((\omega_1 - \delta\omega)t)$. Since we are interested in the nonlinear terms $\beta q_1^3$, we substitute the assumed displacement into the nonlinear term to obtain the following result:

$$\begin{aligned}
\beta q_1^3 &= \beta \left[ X_1 \cos((\omega_1 + \delta\omega)t) + X_2 \cos((\omega_1 - \delta\omega)t) \right]^3 \\
&= \frac{3\beta}{4} \Big[ \left(X_1^3 + 2X_1 X_2^2\right) \cos((\omega_1 + \delta\omega)t) + \left(X_2^3 + 2X_2 X_1^2\right) \cos((\omega_1 - \delta\omega)t) \\
&\quad + X_1^2 X_2 \cos((\omega_1 + 3\delta\omega)t) + X_1 X_2^2 \cos((\omega_1 - 3\delta\omega)t) \\
&\quad + \frac{1}{3} X_1^3 \cos((3\omega_1 + 3\delta\omega)t) + \frac{1}{3} X_2^3 \cos((3\omega_1 - 3\delta\omega)t) \\
&\quad + X_1^2 X_2 \cos((3\omega_1 + \delta\omega)t) + X_1 X_2^2 \cos((3\omega_1 - \delta\omega)t) \Big]
\end{aligned} \quad (S1)$$

From this derivation, it can be seen that the nonlinear term has an effect at and around the drive frequencies $\omega_1 \pm \delta\omega$ and at and around the third harmonic $3\omega_1 \pm \delta\omega$. This means by measuring at either or both of these frequencies, an estimation can be made for the Duffing term.

A similar approach can be taken for coupling terms. In the case of the coupling terms responsible for 1:3 internal resonance between mode 1 and mode 3, $b_{113}^{(1)} q_1^2 q_3$ and $b_{111}^{(3)} q_1^3$, we assume the displacement of each mode to be $q_1(t) = X_1 \cos((\omega_1 + \delta\omega)t) + X_2 \cos((\omega_1 - \delta\omega)t)$ and $q_3(t) = X_3 \cos(\omega_3 t)$. Note that ideally $q_1$ and $q_3$ are both driven at all 3 frequencies, but since these off-resonance drives have little effect compared to the on-resonance ones, only the on-resonance driving frequencies are taken into account. Substitution leads to the following two equations:

$$\begin{aligned}
b_{113}^{(1)} q_1^2 q_3 &= b_{113}^{(1)} \left[ X_1 \cos((\omega_1 + \delta\omega)t) + X_2 \cos((\omega_1 - \delta\omega)t) \right]^2 X_3 \cos(\omega_3 t) \\
&= \frac{b_{113}^{(1)}}{2} \Big[ \left(X_1^2 + X_2^2\right) X_3 \cos(\omega_3 t) + X_1 X_2 X_3 \cos((\omega_3 + 2\delta\omega)t) + X_1 X_2 X_3 \cos((\omega_3 - 2\delta\omega)t) \\
&\quad + X_1 X_2 X_3 \cos((2\omega_1 + \omega_3)t) + X_1 X_2 X_3 \cos((2\omega_1 - \omega_3)t) \\
&\quad + \frac{1}{2} X_1^2 X_3 \cos((2\omega_1 + \omega_3 + 2\delta\omega)t) + \frac{1}{2} X_2^2 X_3 \cos((2\omega_1 + \omega_3 - 2\delta\omega)t) \\
&\quad + \frac{1}{2} X_1^2 X_3 \cos((2\omega_1 - \omega_3 + 2\delta\omega)t) + \frac{1}{2} X_2^2 X_3 \cos((2\omega_1 - \omega_3 - 2\delta\omega)t) \Big]
\end{aligned} \quad (S2)$$

$$\begin{aligned}
b_{111}^{(3)} q_1^3 &= b_{111}^{(3)} \left[ X_1 \cos((\omega_1 + \delta\omega)t) + X_2 \cos((\omega_1 - \delta\omega)t) \right]^3 \\
&= \frac{3 b_{111}^{(3)}}{4} \Big[ \left(X_1^3 + 2X_1 X_2^2\right) \cos((\omega_1 + \delta\omega)t) + \left(X_2^3 + 2X_2 X_1^2\right) \cos((\omega_1 - \delta\omega)t) \\
&\quad + X_1^2 X_2 \cos((\omega_1 + 3\delta\omega)t) + X_1 X_2^2 \cos((\omega_1 - 3\delta\omega)t) \\
&\quad + \frac{1}{3} X_1^3 \cos((3\omega_1 + 3\delta\omega)t) + \frac{1}{3} X_2^3 \cos((3\omega_1 - 3\delta\omega)t) \\
&\quad + X_1^2 X_2 \cos((3\omega_1 + \delta\omega)t) + X_1 X_2^2 \cos((3\omega_1 - \delta\omega)t) \Big]
\end{aligned} \quad (S3)$$

From these equations, it can be seen that the mode 1 term $b_{113}^{(1)}$ does not have an effect at the driving frequencies close to the resonance frequency of mode 1, $\omega_1 \pm \delta\omega$. Similarly, the mode 3 $b_{111}^{(3)}$ term does not have an effect at $\omega_3$. This means measuring only at and around these frequencies will not allow for estimation of these terms since the amplitude of the generated response is too small to be seen below the response of the other driven mode. Therefore, it would be required to measure at additional frequencies like $2\omega_1 + \omega_3$ or any other frequency that can be seen in Eq. S2.

| Nonlinear term | $b_{122}^{(1)} q_1 q_2^2$ | $b_{112}^{(2)} q_1^2 q_2$ |
|---|---|---|
| Close to $\omega_1$ | $\omega_1 \pm \delta, \omega_1 \pm 3\delta, \omega_1 \pm 5\delta, \ldots$ | - |
| Close to $\omega_2$ | - | $\omega_2, \omega_2 \pm 2\delta, \omega_2 \pm 4\delta, \ldots$ |
| Not on resonance | $2\omega_2 \pm \omega_1 \pm \delta, 2\omega_2 \pm \omega_1 \pm 3\delta, 2\omega_2 \pm \omega_1 \pm 5\delta, \ldots$ | $\omega_2 \pm 2\omega_1, \omega_2 \pm 2\omega_1 \pm 2\delta, \omega_2 \pm 2\omega_1 \pm 4\delta, \ldots$ |

TABLE S1. Frequencies where response is generated by dispersive coupling terms, when driven at $\omega_1 \pm \delta$ and $\omega_2$.

Looking at Eq. S3, it can be seen that effects do appear around $\omega_1$. Theoretically, this would allow for the identification of $b_{111}^{(3)}$ by measuring only around the resonance frequencies. However, since this term is part of the mode 3 equation of motion, the motion around $\omega_1$ is far off-resonance. This will cause these effects to be hidden below the mode 1 displacement, unless the modal displacements can be measured separately or accurate modal decomposition can be applied. Therefore, measurement at additional frequencies is also required for the identification of this term.

When repeating this process for coupling terms of the form $\gamma_{i,j} q_i q_j^2$ like the terms seen in the main text, it can be found that these terms do have an effect at the driven frequencies. Specifically, when driving at $\omega_1 \pm \delta$ and $\omega_2$, the frequencies that are obtained following the analysis of $b_{122}^{(1)} q_1 q_2^2$ and $b_{112}^{(2)} q_1^2 q_2$ can be seen in Tab. S1, which clearly shows these terms generate sidebands around the resonance frequency of their corresponding equation of motion.

## SUPPLEMENTARY NOTE III. DERIVATION OF FREQUENCY DOMAIN EQUATIONS OF MOTION

We start with the nonlinear equations of motion of $N$ coupled degrees-of-freedom as follows:

$$\ddot{q}_n + \frac{\omega_n}{Q_n} \dot{q}_n + \omega_n^2 q_n + \sum_{i=1}^{N} \sum_{j=1}^{N} \sum_{k=1}^{N} b_{ijk}^{(n)} q_i q_j q_k = \sum_{d=1}^{D} F_d^{(n)} \cos(\Omega_d t + \phi_d), \qquad n = 1, \ldots, N, \tag{S4}$$

where $q_n$ describe the modal displacements, $\omega_n$ and $Q_n$ are the linear angular frequency and Q-factor of mode $n$ and the parameters $b_{ijk}^{(n)}$ represent cubic stiffness terms. The force $F_d^{(n)}$ represents the multi-tone drives placed at frequencies $\Omega_d$ at or in the spectral neighbourhood of the resonance frequencies. Here, we derive the equations seen in Eq. 2 in the main text, following the assumed form of the solution $q_n(t) = \sum_{r=1}^{R} \hat{q}_n^{(r)} e^{j\Omega_r t}$, where $\hat{q}_n^{(r)} = A_r e^{i\phi_r}$ are the complex amplitudes at angular frequencies $\Omega_r$. Based on this solution, the velocity and acceleration terms can be written as:

$$\dot{q}_n(t) = \sum_{r=1}^{R} j\Omega_r \hat{q}_n^{(r)} e^{j\Omega_r t}, \tag{S5}$$

$$\ddot{q}_n(t) = \sum_{r=1}^{R} -\Omega_r^2 \hat{q}_n^{(r)} e^{j\Omega_r t}. \tag{S6}$$

The forcing terms can be evaluated by splitting each term in a sine and cosine term:

$$\sum_{d=1}^{D} F_d^{(n)} \cos(\Omega_d t + \phi_d) = \sum_{d=1}^{D} F_{C,d}^{(n)} \cos(\Omega_d t) + F_{S,d}^{(n)} \sin(\Omega_d t), \tag{S7}$$

where $F_{C,d}^{(n)}$ and $F_{S,d}^{(n)}$ are the in-and out-of-phase force components, such that $F_d^{(n)^2} = F_{C,d}^{(n)^2} + F_{S,d}^{(n)^2}$ and $\phi_d = \arctan\left(F_{S,d}^{(n)}/F_{C,d}^{(n)}\right)$. Based on this, we define the $\Omega_r$-component of the sine and cosine terms as:

$$\hat{f}_{C,d}^{(r)} = \mathcal{F}^{(r)}\{\cos(\Omega_d t)\}, \qquad \hat{f}_{S,d}^{(r)} = \mathcal{F}^{(r)}\{\sin(\Omega_d t)\}. \tag{S8}$$

Finally, the nonlinear terms can be evaluated at frequencies $\Omega_r$. For this, constant $\hat{g}_{ijk}^{(r)}$ is defined as follows:

$$\hat{g}_{ijk}^{(r)} = \mathcal{F}^{(r)}\{q_i q_j q_k\} = \mathcal{F}^{(r)}\left\{\sum_a \hat{q}_i^{(a)} e^{i\Omega_a t} \sum_b \hat{q}_j^{(b)} e^{i\Omega_b t} \sum_c \hat{q}_k^{(c)} e^{i\Omega_c t}\right\}. \quad (S9)$$

$q_n = \sum_r \hat{q}_n^{(r)} e^{i\Omega_r t}$ for $n = i, j, k$, from the measured complex modal amplitudes. Then, by multiplying the displacements and performing a Fourier transform at frequency $\Omega_r$, such that $\hat{g}_{ijk}^{(r)} = \mathcal{F}^{(r)}\{q_i q_j q_k\}$.

Here $\mathcal{F}^{(r)}$ represents the Fourier transform evaluated at frequency $\Omega_r$ and $*$ denotes convolution in the frequency domain. The term $\hat{g}_{ijk}^{(r)}$ can be seen as the complex amplitude of the nonlinear term at frequency $\Omega_r$.

By inserting all above definitions into S4, we obtain the following equation:

$$\sum_{r=1}^{R} -\Omega_r^2 \hat{q}_n^{(r)} e^{j\Omega_r t} + \frac{\omega_n}{Q_n} j\Omega_r \hat{q}_n^{(r)} e^{j\Omega_r t} + \omega_n^2 \hat{q}_n^{(r)} e^{j\Omega_r t} + \sum_{i=1}^{N}\sum_{j=1}^{N}\sum_{k=1}^{N} b_{ijk}^{(n)} \hat{g}_{ijk}^{(r)} e^{j\Omega_r t}$$
$$= \sum_{r=1}^{R}\sum_{d=1}^{D} F_{C,d}^{(n)} \hat{f}_{C,d}^{(r)} e^{j\Omega_r t} + F_{S,d}^{(n)} \hat{f}_{S,d}^{(r)} e^{j\Omega_r t}, \quad n = 1, ..., N. \quad (S10)$$

This equation is split into $R$ separate equations for each sideband frequency, leading to Eq. 2 in the main text:

$$\Omega_r^2 \hat{q}_n^{(r)} + \frac{\omega_n}{Q_n} j\Omega_r \hat{q}_n^{(r)} + \omega_n^2 \hat{q}_n^{(r)} + \sum_{i=1}^{N}\sum_{j=1}^{N}\sum_{k=1}^{N} b_{ijk}^{(n)} \hat{g}_{ijk}^{(r)} = \sum_{d=1}^{D} F_{C,d}^{(n)} \hat{f}_{C,d}^{(r)} + F_{S,d}^{(n)} \hat{f}_{S,d}^{(r)}, \quad r = 1, ..., R, \quad n = 1, ..., N. \quad (S11)$$

### SUPPLEMENTARY NOTE VI. PARAMETER QUANTIFICATION EQUATIONS IN MATRIX FORM

In multimode excitation, the response can be governed by linear dynamics and single-mode Duffing nonlinearities, whose effects can overshadow the comparatively weaker signatures of intermodal coupling. As a result, simultaneous fitting reduces the sensitivity of the reconstruction to coupling terms and can lead to inaccurate estimates. To overcome this, we adopt a sequential identification strategy in which single-mode parameters are determined first and the obtained estimates are used when quantifying the coupling strengths.

To obtain the single-mode parameters, two-tone response is measured for each individual mode. Using the obtained complex amplitudes, Eq. 3 from the main text can be constructed in the following matrix form, here shown for the first mode:

$$\underbrace{\begin{bmatrix} j\Omega_1 \hat{q}_1^{(1)} & \hat{q}_1^{(1)} & \hat{g}_{111}^{(1)} & \hat{f}_{C,1}^{(1)} & \hat{f}_{S,1}^{(1)} & \hat{f}_{C,2}^{(1)} & \hat{f}_{S,2}^{(1)} \\ j\Omega_2 \hat{q}_1^{(2)} & \hat{q}_1^{(2)} & \hat{g}_{111}^{(2)} & \hat{f}_{C,1}^{(2)} & \hat{f}_{S,1}^{(2)} & \hat{f}_{C,2}^{(2)} & \hat{f}_{S,2}^{(2)} \\ \vdots & \vdots & \vdots & \vdots & \vdots & \vdots & \vdots \\ j\Omega_R \hat{q}_1^{(R)} & \hat{q}_1^{(R)} & \hat{g}_{111}^{(R)} & \hat{f}_{C,1}^{(R)} & \hat{f}_{S,1}^{(R)} & \hat{f}_{C,2}^{(R)} & \hat{f}_{S,2}^{(R)} \end{bmatrix}}_{H} \underbrace{\begin{bmatrix} \frac{\omega_1}{Q_1} \\ \omega_1^2 \\ b_{111}^{(1)} \\ -F_{C,1}^{(1)} \\ -F_{S,1}^{(1)} \\ -F_{C,2}^{(1)} \\ -F_{S,2}^{(1)} \end{bmatrix}}_{\bar{p}} = -\underbrace{\begin{bmatrix} -\Omega_1^2 \hat{q}_1^{(1)} \\ -\Omega_2^2 \hat{q}_1^{(2)} \\ \vdots \\ -\Omega_R^2 \hat{q}_1^{(R)} \end{bmatrix}}_{\bar{Q}}. \quad (S12)$$

Solving for $\bar{p}$, we obtain the linear resonance $\omega_1$, Q-factor $Q_1$ and Duffing term $b_{111}^{(1)}$. Repeating this for other modes gives all linear and Duffing parameters for higher-order modes.

Next, to estimate the coupling parameters, three driving tones are applied following the procedure described in the main text. To estimate cubic couplings between modes 1 and 2, the Eq. 4 in matrix form is then constructed:

$$\underbrace{\begin{bmatrix} j\Omega_1 \hat{q}_1^{(1)} & \hat{q}_1^{(1)} & \hat{g}_{111}^{(1)} & \hat{g}_{122}^{(1)} & \hat{f}_{C,1}^{(1)} & \hat{f}_{S,1}^{(1)} & \cdots & \hat{f}_{C,D}^{(1)} & \hat{f}_{S,D}^{(1)} \\ j\Omega_2 \hat{q}_1^{(2)} & \hat{q}_1^{(2)} & \hat{g}_{111}^{(2)} & \hat{g}_{122}^{(2)} & \hat{f}_{C,1}^{(2)} & \hat{f}_{S,1}^{(2)} & \cdots & \hat{f}_{C,D}^{(2)} & \hat{f}_{S,D}^{(2)} \\ \vdots & \vdots & \vdots & \vdots & \vdots & \vdots & \ddots & \vdots & \vdots \\ j\Omega_R \hat{q}_1^{(R)} & \hat{q}_1^{(R)} & \hat{g}_{111}^{(R)} & \hat{g}_{122}^{(R)} & \hat{f}_{C,1}^{(R)} & \hat{f}_{S,1}^{(R)} & \cdots & \hat{f}_{C,3}^{(R)} & \hat{f}_{S,3}^{(R)} \end{bmatrix}}_{\boldsymbol{H}} \underbrace{\begin{bmatrix} \frac{\omega_1}{Q_1} \\ \omega_1^2 \\ b_{111}^{(1)} \\ b_{122}^{(1)} \\ -F_{C,1}^{(1)} \\ -F_{S,1}^{(1)} \\ \vdots \\ -F_{C,3}^{(1)} \\ -F_{S,3}^{(1)} \end{bmatrix}}_{\bar{\boldsymbol{p}}} = -\underbrace{\begin{bmatrix} -\Omega_1^2 \hat{q}_1^{(1)} \\ -\Omega_2^2 \hat{q}_1^{(2)} \\ \vdots \\ -\Omega_R^2 \hat{q}_1^{(R)} \end{bmatrix}}_{\bar{\boldsymbol{Q}}}, \tag{S13}$$

where we consider all Duffing and linear parameters to be known following the identification performed with a two-tone drive. Since all parameters in $\bar{p}$ should be unknown, the damping, stiffness and Duffing terms from $H$ and $\bar{p}$ are then moved to $\bar{Q}$. This results in a much simpler equation:

$$\underbrace{\begin{bmatrix} \hat{g}_{122}^{(1)} & \hat{f}_{C,1}^{(1)} & \hat{f}_{S,1}^{(1)} & \cdots & \hat{f}_{C,3}^{(1)} & \hat{f}_{S,3}^{(1)} \\ \hat{g}_{122}^{(2)} & \hat{f}_{C,1}^{(2)} & \hat{f}_{S,1}^{(2)} & \cdots & \hat{f}_{C,3}^{(2)} & \hat{f}_{S,3}^{(2)} \\ \vdots & \vdots & \vdots & \ddots & \vdots & \vdots \\ \hat{g}_{122}^{(R)} & \hat{f}_{C,1}^{(R)} & \hat{f}_{S,1}^{(R)} & \cdots & \hat{f}_{C,3}^{(R)} & \hat{f}_{S,3}^{(R)} \end{bmatrix}}_{\boldsymbol{H}} \underbrace{\begin{bmatrix} b_{122}^{(1)} \\ -F_{C,1}^{(1)} \\ -F_{S,1}^{(1)} \\ \vdots \\ -F_{C,3}^{(1)} \\ -F_{S,3}^{(1)} \end{bmatrix}}_{\bar{\boldsymbol{p}}} = -\underbrace{\begin{bmatrix} -\Omega_1^2 \hat{q}_1^{(1)} + j\Omega_1 \frac{\omega_1}{Q_1} \hat{q}_1^{(1)} + \omega_1^2 \hat{q}_1^{(1)} + b_{111}^{(1)} \hat{g}_{111}^{(1)} \\ -\Omega_2^2 \hat{q}_1^{(2)} + j\Omega_2 \frac{\omega_1}{Q_1} \hat{q}_1^{(2)} + \omega_1^2 \hat{q}_1^{(2)} + b_{111}^{(1)} \hat{g}_{111}^{(2)} \\ \vdots \\ -\Omega_R^2 \hat{q}_1^{(R)} + j\Omega_R \frac{\omega_1}{Q_1} \hat{q}_1^{(R)} + \omega_1^2 \hat{q}_1^{(R)} + b_{111}^{(1)} \hat{g}_{111}^{(R)} \end{bmatrix}}_{\bar{\boldsymbol{Q}}}. \tag{S14}$$

.

Here, only the coupling term $b_{122}^{(1)}$ and driving forces are obtained in a least squares optimization, leading to a more accurate result. Naturally, this procedure can be applied when more coupling parameters are considered or when looking at higher-order nonlinear terms.

Adding to the above simplification, we can take into account the nonlinear potential linking together certain coupling terms. For the coupling terms between the first two modes which are considered in the main text, the potential takes the form $U_{\mathrm{nl}} = \frac{1}{2}\tilde{\gamma}q_1^2 q_2^2$. Since terms in the equation of motion follow from the derivative of $U_{\mathrm{nl}}$ with respect to generalized coordinates $q_1$ and $q_2$, it can be found that $b_{122}^{(1)} = b_{112}^{(2)} = \tilde{\gamma}$. Therefore, instead of estimating $b_{122}^{(1)}$ and $b_{112}^{(2)}$ separately, we obtain $\tilde{\gamma}$ directly. This is done by taking Eq. S14 for each mode, and combining them in an equation of the form:

$$\begin{bmatrix} \boldsymbol{H}_{1,\text{uncoupled}} & 0 & \boldsymbol{H}_{1,\text{coupled}} \\ 0 & \boldsymbol{H}_{2,\text{uncoupled}} & \boldsymbol{H}_{2,\text{coupled}} \end{bmatrix} \begin{bmatrix} \bar{\boldsymbol{p}}_{1,\text{uncoupled}} \\ \bar{\boldsymbol{p}}_{2,\text{uncoupled}} \\ \bar{\boldsymbol{p}}_{\text{coupled}} \end{bmatrix} = -\begin{bmatrix} \bar{\boldsymbol{Q}}_1 \\ \bar{\boldsymbol{Q}}_2 \end{bmatrix} \tag{S15}$$

Where $H_{i,\text{uncoupled}}$ represents the part of $H_i$ that is not coupled via the potential and $H_{i,\text{coupled}}$ represents the coupled part that is linked via the potential. Within this coupled part the matrix, we effectively add the relation between couplings from the nonlinear potential as a constraint. Similar for $\bar{p}_{i,\text{uncoupled}}$, but $\bar{p}_{i,\text{coupled}}$ consists of the potential coefficients of the coupled terms, which for dispersive coupling would be $\tilde{\gamma}$. Evaluating this matrix for the coupling analysis between mode 1 and mode 2, taking into account previously estimated results leads to the following equation:

$$\underbrace{\begin{bmatrix}
\hat{f}_{\mathrm{C},1}^{(1)} & \hat{f}_{\mathrm{S},1}^{(1)} & \cdots & \hat{f}_{\mathrm{C},D}^{(1)} & \hat{f}_{\mathrm{S},D}^{(1)} & & \cdots & 0 & \cdots & & \hat{g}_{122}^{(1)} \\
\hat{f}_{\mathrm{C},1}^{(2)} & \hat{f}_{\mathrm{S},1}^{(2)} & \cdots & \hat{f}_{\mathrm{C},D}^{(2)} & \hat{f}_{\mathrm{S},D}^{(2)} & & \cdots & 0 & \cdots & & \hat{g}_{122}^{(2)} \\
\vdots & \vdots & \ddots & \vdots & \vdots & & & \vdots & & & \vdots \\
\hat{f}_{\mathrm{C},1}^{(R)} & \hat{f}_{\mathrm{S},1}^{(R)} & \cdots & \hat{f}_{\mathrm{C},D}^{(R)} & \hat{f}_{\mathrm{S},D}^{(R)} & & \cdots & 0 & \cdots & & \hat{g}_{122}^{(R)} \\
& \cdots & 0 & \cdots & & \hat{f}_{\mathrm{C},1}^{(1)} & \hat{f}_{\mathrm{S},1}^{(1)} & \cdots & \hat{f}_{\mathrm{C},D}^{(1)} & \hat{f}_{\mathrm{S},D}^{(1)} & \hat{g}_{112}^{(1)} \\
& \cdots & 0 & \cdots & & \hat{f}_{\mathrm{C},1}^{(2)} & \hat{f}_{\mathrm{S},1}^{(2)} & \cdots & \hat{f}_{\mathrm{C},D}^{(2)} & \hat{f}_{\mathrm{S},D}^{(2)} & \hat{g}_{112}^{(2)} \\
& \vdots & \vdots & & & \vdots & \vdots & \ddots & \vdots & \vdots & \vdots \\
& \cdots & 0 & \cdots & & \hat{f}_{\mathrm{C},1}^{(R)} & \hat{f}_{\mathrm{S},1}^{(R)} & \cdots & \hat{f}_{\mathrm{C},D}^{(R)} & \hat{f}_{\mathrm{S},D}^{(R)} & \hat{g}_{112}^{(R)}
\end{bmatrix}}_{\boldsymbol{H}_{\mathrm{combined}}} \underbrace{\begin{bmatrix} -F_{\mathrm{C},1}^{(1)} \\ -F_{\mathrm{S},1}^{(1)} \\ \vdots \\ -F_{\mathrm{C},D}^{(1)} \\ -F_{\mathrm{S},D}^{(1)} \\ -F_{\mathrm{C},1}^{(2)} \\ -F_{\mathrm{S},1}^{(2)} \\ \vdots \\ -F_{\mathrm{C},D}^{(2)} \\ -F_{\mathrm{S},D}^{(2)} \\ \tilde{\gamma} \end{bmatrix}}_{\bar{\boldsymbol{p}}_{\mathrm{combined}}} =$$

$$-\underbrace{\begin{bmatrix}
-\Omega_1^2 \hat{q}_1^{(1)} + j\Omega_1 \frac{\omega_1}{Q_1} \hat{q}_1^{(1)} + \omega_1^2 \hat{q}_1^{(1)} + b_{111}^{(1)} \hat{g}_{111}^{(1)} \\
-\Omega_2^2 \hat{q}_1^{(2)} + j\Omega_2 \frac{\omega_1}{Q_1} \hat{q}_1^{(2)} + \omega_1^2 \hat{q}_1^{(2)} + b_{111}^{(1)} \hat{g}_{111}^{(2)} \\
\vdots \\
-\Omega_R^2 \hat{q}_1^{(R)} + j\Omega_R \frac{\omega_1}{Q_1} \hat{q}_1^{(R)} + \omega_1^2 \hat{q}_1^{(R)} + b_{111}^{(1)} \hat{g}_{111}^{(R)} \\
-\Omega_1^2 \hat{q}_2^{(1)} + j\Omega_1 \frac{\omega_2}{Q_2} \hat{q}_2^{(1)} + \omega_2^2 \hat{q}_2^{(1)} + b_{222}^{(2)} \hat{g}_{222}^{(1)} \\
-\Omega_2^2 \hat{q}_2^{(2)} + j\Omega_2 \frac{\omega_2}{Q_2} \hat{q}_2^{(2)} + \omega_2^2 \hat{q}_2^{(2)} + b_{222}^{(2)} \hat{g}_{222}^{(2)} \\
\vdots \\
-\Omega_R^2 \hat{q}_2^{(R)} + j\Omega_R \frac{\omega_2}{Q_2} \hat{q}_2^{(R)} + \omega_2^2 \hat{q}_2^{(R)} + b_{222}^{(2)} \hat{g}_{222}^{(R)}
\end{bmatrix}}_{\bar{\boldsymbol{Q}}_{\mathrm{combined}}} \quad (S16)$$

By solving this equation for $\bar{p}_{\mathrm{combined}}$, the coupling parameters for the equations of motion can be calculated based on the nonlinear potential parameter $\tilde{\gamma}$. This procedure can naturally also be applied on other coupling terms, including multi-modal coupling, to ensure the fitted model will be physically meaningful.

## SUPPLEMENTARY NOTE V. NUMERICAL ROBUSTNESS TO NOISE

To analyse the robustness of our parameter estimation approach to noise, numerical simulations are performed based on experiments on a sample with $L_s = 50\,\mu\mathrm{m}$. The parameters used to define the equations of motion can be seen in Table. S2. Drive frequencies are placed at a spacing $df$ of 20 Hz.

TABLE S2. Mass-normalized parameters used in simulations

| Mode number $(i)$ | 1 | 2 |
|---|---|---|
| $f_i$ | $207.582 \times 10^3$ | $415.842 \times 10^3$ |
| $Q_i$ | 116232 | 599173 |
| $F_1^{(i)} = F_2^{(i)}$ | 10.06 | 1.81 |
| $\beta_i$ | $2.68 \times 10^{22}$ | $3.70 \times 10^{23}$ |
| $\gamma_{1,2}$ | $8.57 \times 10^{22}$ | $8.57 \times 10^{22}$ |

Simulations are performed in Python, using the function scipy.integrate.odeint() for $2Q_i/f_i$ oscillations to reach steady state, and then for another 4 slow beats $f_1/df$ to have an integration time of 200 ms, similar to experiments shown in the main text. To mimic the experiments, 3 simulations are carried out for single mode identification of mode 1, single mode identification of mode 2 and coupling identification between the two modes. A discrete Fourier transform (DFT) is applied on the time domain data to obtain the complex amplitudes at sideband frequencies, which without noise results in data shown in Fig. S2.

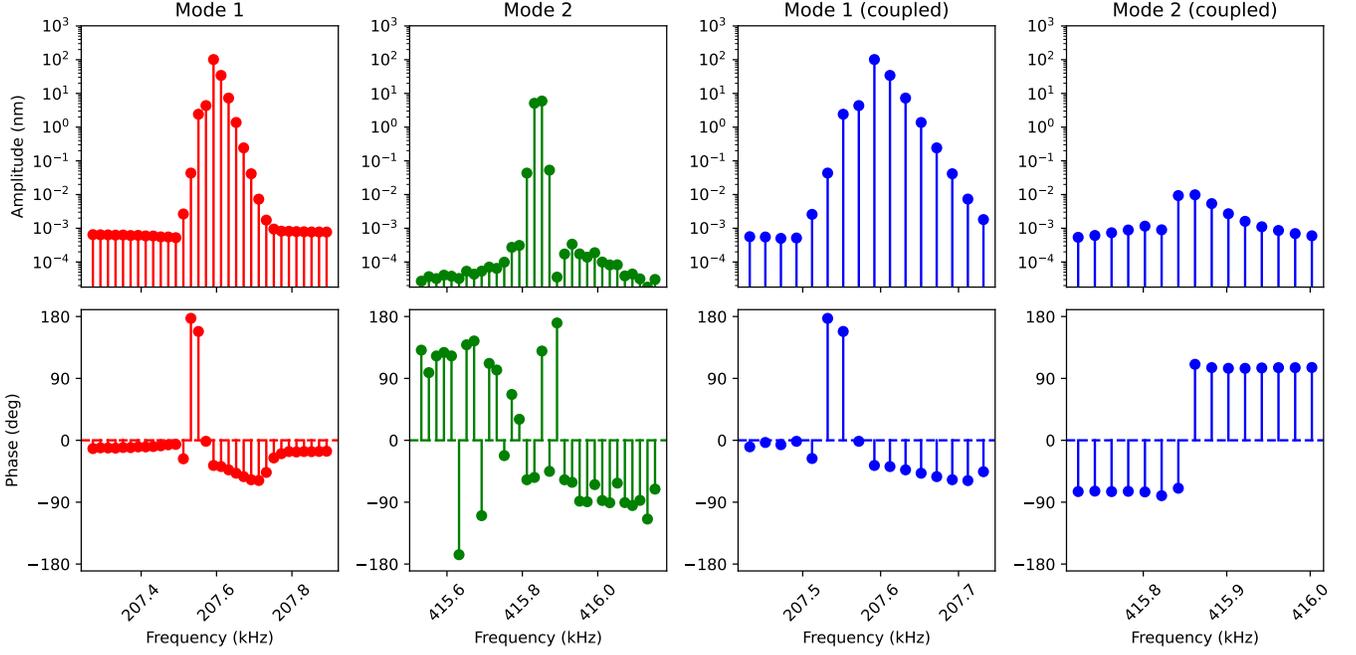

FIG. S2. Simulated intermodulation response for single mode identification (red and green) and coupling identification (blue).

To analyse the effects of noise, a random signal with a given RMS amplitude is added onto the time domain signal. A DFT is again applied to the noisy signal, and the resulting intermodulation products are used to quantify the parameters in the system. This is repeated 30 times for each noise level, for RMS amplitudes ranging from $10^{-3}$ nm to $10^2$ nm. The resulting fitted parameters can be seen in Fig. S3.

Since the order of magnitude for the thermomechanical noise in the devices in our experiments is on the order of $10^{-1}$ nm [2], accurate estimation of all terms is expected for experimental results using our multi-tone excitation approach.

## SUPPLEMENTARY NOTE VI. EXPERIMENTALLY IDENTIFIED PARAMETERS AND VALUES USED FOR COMPARISON

To complement Fig 4 and Fig 5 in the main text, this section reports the numerical values of the parameters that are shown or used in those figures. This includes the results obtained from the experimental methodology, as well as data used for comparison from previous studies and simulations.

Table S3 and Table S4 show the experimental results from the two-mode identification seen in Fig 4b-e in the main text, of the methodology proposed in this work and of a previous study respectively [3].

Table S5 and Table S6 show the parameters used to calculate the nonlinear potential shown in Fig. 5, from experiments and simulation respectively. Note that Table S5 shows the same results as the table in Fig. 5 in the main text, but following the definitions of nonlinear terms as shown in Eq. 1.

## SUPPLEMENTARY NOTE VII. SAMPLE FABRICATION

Devices have been fabricated from high-stress $Si_3N_4$ layers by electron beam lithography and reactive ion etching. The layers have been grown with a thickness of 90 nm on a silicon substrate using low-pressure chemical vapour deposition. The devices are then suspended by a fluorine-based ($SF_6$) deep reactive ion underetching step. After fabrication, the nano-strings have an initial isotropic stress $\sigma_0 = 1.06$ GPa, Young's modulus $E = 271$ GPa, Poisson's ratio $\nu = 0.23$, mass density $\rho = 3100$ kg m$^{-3}$ and an intrinsic Q-factor $Q_0 = 4527$. See previous work for more details [2–4].

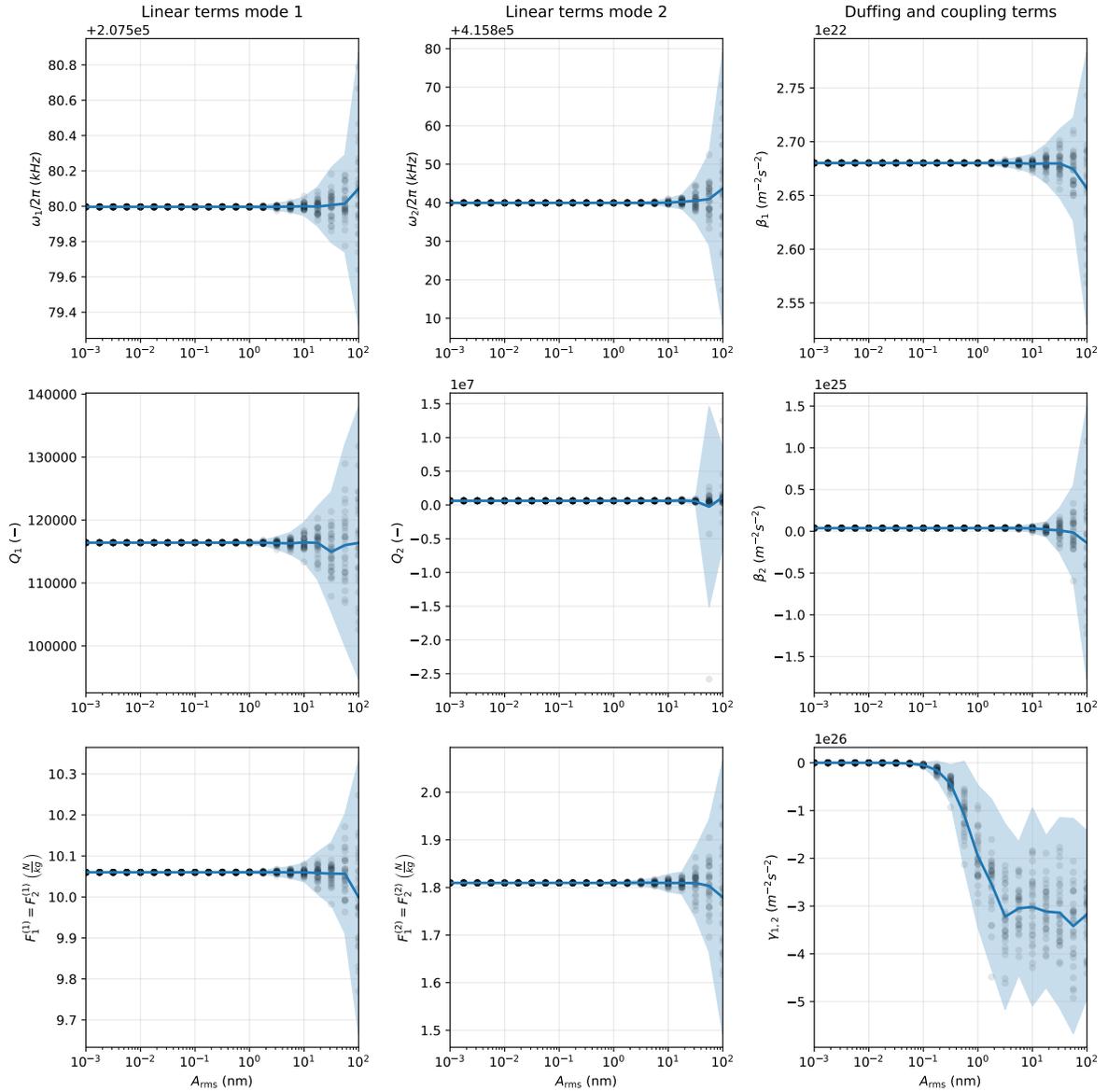

FIG. S3. Estimated parameters at different noise levels. For each level, random noise is added to the time-response 30 times. After addition of noise, a DFT is performed to obtain the intermodulation products, simulating the effect of the multi-frequency lock-in amplifier in the experimental setup. Using the noisy frequency domain, estimation is performed. Dots represent identified values, the blue line shows the average across noise levels, the blue shaded area shows $\pm 3\sigma$.

TABLE S3. Experimentally obtained parameters for the first two modes of devices with different support length $L_s$. Complementary to the mean of the values seen in Fig. 4b-d and the mean of the errorbars seen in Fig 4e in the main text.

| Support length ($\mu m$) | 30 | 50 | 70 | 90 |
|---|---|---|---|---|
| $f_1$ (kHz) | 345.880 | 207.669 | 155.762 | 126.770 |
| $Q_1$ | 123418.8 | 144688.5 | 75958.5 | 264604.7 |
| $F_1^{(1)} = F_2^{(1)}$ (N/kg) | 6.1 | 2.7 | 2.1 | 1.3 |
| $f_2$ (kHz) | 691.942 | 416.031 | 312.602 | 255.072 |
| $Q_2$ | 226686.9 | 379493.0 | 103812.4 | 124812.9 |
| $F_1^{(2)} = F_2^{(2)}$ (N/kg) | 2.2 | 2.1 | 5.0 | 2.2 |
| $\beta_1$ (m$^{-2}$s$^{-2}$) | $8.00 \times 10^{22}$ | $2.93 \times 10^{22}$ | $1.65 \times 10^{22}$ | $2.64 \times 10^{22}$ |
| $\beta_2$ (m$^{-2}$s$^{-2}$) | $1.23 \times 10^{24}$ | $4.64 \times 10^{23}$ | $3.09 \times 10^{23}$ | $1.40 \times 10^{23}$ |
| $\gamma_{1,2}$ (m$^{-2}$s$^{-2}$) | $2.59 \times 10^{23}$ | $2.89 \times 10^{23}$ | $1.08 \times 10^{23}$ | $8.06 \times 10^{22}$ |

TABLE S4. Experimentally obtained parameters for the first two modes of devices with different support length $L_s$, by analysing the change in slope of the backbone curve of the first mode [3]. Complementary to the diamonds seen in Fig. 4e in the main text.

| Support length ($\mu m$) | 30 | 50 | 70 | 90 |
|---|---|---|---|---|
| $f_1$ (kHz) | 347.320 | 207.892 | 156.532 | 127.701 |
| $f_2$ (kHz) | 694.724 | 416.427 | 314.299 | 256.128 |
| $\beta_1$ (m$^{-2}$s$^{-2}$) | $7.73 \times 10^{22}$ | $2.01 \times 10^{22}$ | $1.17 \times 10^{22}$ | $8.60 \times 10^{21}$ |
| $\beta_2$ (m$^{-2}$s$^{-2}$) | $1.66 \times 10^{24}$ | $2.92 \times 10^{23}$ | $1.83 \times 10^{23}$ | $1.49 \times 10^{23}$ |
| $\gamma_{1,2}$ (m$^{-2}$s$^{-2}$) | $9.33 \times 10^{22}$ | $6.07 \times 10^{22}$ | $3.44 \times 10^{22}$ | $3.80 \times 10^{22}$ |

TABLE S5. Experimentally obtained parameters for the first five modes of the device with support length $L_s = 50\mu m$. Complementary Fig. 5 in the main text.

| Mode number ($i$) | 1 | 2 | 3 | 4 | 5 |
|---|---|---|---|---|---|
| $f_i$ (kHz) | 207.582 | 415.842 | 626.926 | 841.449 | 1061.035 |
| $Q_i$ | 116232.6 | 599173.3 | 41181.4 | 140505.3 | 128064.5 |
| $b_{11i}^{(i)}$ (m$^{-2}$s$^{-2}$) | $2.68 \times 10^{22}$ | $8.57 \times 10^{22}$ | $2.51 \times 10^{23}$ | $4.13 \times 10^{23}$ | $6.24 \times 10^{23}$ |
| $b_{22i}^{(i)}$ (m$^{-2}$s$^{-2}$) | $8.57 \times 10^{22}$ | $3.70 \times 10^{23}$ | $6.97 \times 10^{23}$ | $2.33 \times 10^{24}$ | $2.66 \times 10^{24}$ |
| $b_{33i}^{(i)}$ (m$^{-2}$s$^{-2}$) | $2.51 \times 10^{23}$ | $6.97 \times 10^{23}$ | $2.16 \times 10^{24}$ | $2.91 \times 10^{24}$ | $5.40 \times 10^{24}$ |
| $b_{44i}^{(i)}$ (m$^{-2}$s$^{-2}$) | $4.13 \times 10^{23}$ | $2.33 \times 10^{24}$ | $2.91 \times 10^{24}$ | $6.27 \times 10^{24}$ | $5.47 \times 10^{24}$ |
| $b_{55i}^{(i)}$ (m$^{-2}$s$^{-2}$) | $6.24 \times 10^{23}$ | $2.66 \times 10^{24}$ | $5.40 \times 10^{24}$ | $5.47 \times 10^{24}$ | $1.57 \times 10^{25}$ |

TABLE S6. Simulated parameters for the first five modes of the device with support length $L_s = 50\mu m$, taken from [3]. Complementary to Fig. 4f in the main text.

| Mode number ($i$) | 1 | 2 | 3 | 4 | 5 |
|---|---|---|---|---|---|
| $f_i$ (Hz) | $2.11 \times 10^5$ | $4.24 \times 10^5$ | $6.39 \times 10^5$ | $8.58 \times 10^5$ | $1.08 \times 10^6$ |
| $Q_i$ | $4.28 \times 10^5$ | $2.76 \times 10^5$ | $1.75 \times 10^5$ | $1.16 \times 10^5$ | $8.22 \times 10^4$ |
| $b_{111}^{(i)}$ (m$^{-2}$s$^{-2}$) | $3.56 \times 10^{22}$ | $2.08 \times 10^{18}$ | $-1.66 \times 10^{21}$ | $4.85 \times 10^{18}$ | $-3.35 \times 10^{21}$ |
| $b_{112}^{(i)}$ (m$^{-2}$s$^{-2}$) | $6.64 \times 10^{18}$ | $1.49 \times 10^{23}$ | $1.19 \times 10^{19}$ | $-1.88 \times 10^{21}$ | $1.78 \times 10^{19}$ |
| $b_{113}^{(i)}$ (m$^{-2}$s$^{-2}$) | $-5.12 \times 10^{21}$ | $1.15 \times 10^{19}$ | $3.29 \times 10^{23}$ | $2.17 \times 10^{19}$ | $-3.25 \times 10^{21}$ |
| $b_{114}^{(i)}$ (m$^{-2}$s$^{-2}$) | $1.36 \times 10^{19}$ | $-1.91 \times 10^{21}$ | $2.16 \times 10^{19}$ | $5.81 \times 10^{23}$ | $3.19 \times 10^{19}$ |
| $b_{115}^{(i)}$ (m$^{-2}$s$^{-2}$) | $-1.08 \times 10^{22}$ | $2.10 \times 10^{19}$ | $-3.23 \times 10^{21}$ | $3.19 \times 10^{19}$ | $9.05 \times 10^{23}$ |
| $b_{122}^{(i)}$ (m$^{-2}$s$^{-2}$) | $1.58 \times 10^{23}$ | $2.57 \times 10^{19}$ | $-2.64 \times 10^{21}$ | $3.46 \times 10^{19}$ | $-5.85 \times 10^{21}$ |
| $b_{123}^{(i)}$ (m$^{-2}$s$^{-2}$) | $5.92 \times 10^{21}$ | $-2.73 \times 10^{22}$ | $-4.98 \times 10^{22}$ | $2.51 \times 10^{22}$ | $6.64 \times 10^{20}$ |
| $b_{124}^{(i)}$ (m$^{-2}$s$^{-2}$) | $-8.68 \times 10^{21}$ | $2.15 \times 10^{22}$ | $2.62 \times 10^{22}$ | $-8.59 \times 10^{22}$ | $4.39 \times 10^{22}$ |
| $b_{125}^{(i)}$ (m$^{-2}$s$^{-2}$) | $-6.38 \times 10^{21}$ | $1.10 \times 10^{22}$ | $-3.03 \times 10^{20}$ | $4.38 \times 10^{22}$ | $-1.34 \times 10^{23}$ |
| $b_{133}^{(i)}$ (m$^{-2}$s$^{-2}$) | $3.51 \times 10^{23}$ | $3.50 \times 10^{19}$ | $-5.54 \times 10^{22}$ | $7.19 \times 10^{19}$ | $-1.23 \times 10^{22}$ |
| $b_{134}^{(i)}$ (m$^{-2}$s$^{-2}$) | $-7.02 \times 10^{21}$ | $2.38 \times 10^{22}$ | $4.84 \times 10^{22}$ | $-1.4 \times 10^{23}$ | $3.16 \times 10^{21}$ |
| $b_{135}^{(i)}$ (m$^{-2}$s$^{-2}$) | $-1.05 \times 10^{22}$ | $9.62 \times 10^{19}$ | $2.24 \times 10^{22}$ | $2.05 \times 10^{20}$ | $-2.1 \times 10^{23}$ |
| $b_{144}^{(i)}$ (m$^{-2}$s$^{-2}$) | $6.22 \times 10^{23}$ | $6.41 \times 10^{19}$ | $-2.73 \times 10^{22}$ | $2.28 \times 10^{20}$ | $-5.09 \times 10^{22}$ |
| $b_{145}^{(i)}$ (m$^{-2}$s$^{-2}$) | $3.74 \times 10^{21}$ | $4.17 \times 10^{22}$ | $2.74 \times 10^{21}$ | $-2.39 \times 10^{22}$ | $1.24 \times 10^{23}$ |
| $b_{155}^{(i)}$ (m$^{-2}$s$^{-2}$) | $9.70 \times 10^{23}$ | $1.05 \times 10^{20}$ | $-4.04 \times 10^{22}$ | $2.24 \times 10^{20}$ | $-2.78 \times 10^{23}$ |
| $b_{222}^{(i)}$ (m$^{-2}$s$^{-2}$) | $1.05 \times 10^{19}$ | $5.44 \times 10^{23}$ | $3.76 \times 10^{19}$ | $-1.96 \times 10^{22}$ | $4.75 \times 10^{19}$ |
| $b_{223}^{(i)}$ (m$^{-2}$s$^{-2}$) | $-3.56 \times 10^{21}$ | $6.8 \times 10^{19}$ | $1.25 \times 10^{24}$ | $7.71 \times 10^{19}$ | $-3.17 \times 10^{22}$ |
| $b_{224}^{(i)}$ (m$^{-2}$s$^{-2}$) | $3.23 \times 10^{19}$ | $-6.13 \times 10^{22}$ | $7.79 \times 10^{19}$ | $2.21 \times 10^{24}$ | $1.31 \times 10^{20}$ |
| $b_{225}^{(i)}$ (m$^{-2}$s$^{-2}$) | $-6.13 \times 10^{21}$ | $1.17 \times 10^{20}$ | $-3.13 \times 10^{22}$ | $1.3 \times 10^{20}$ | $3.44 \times 10^{24}$ |
| $b_{233}^{(i)}$ (m$^{-2}$s$^{-2}$) | $3.38 \times 10^{19}$ | $1.26 \times 10^{24}$ | $1.50 \times 10^{20}$ | $-2.21 \times 10^{22}$ | $1.59 \times 10^{20}$ |
| $b_{234}^{(i)}$ (m$^{-2}$s$^{-2}$) | $1.99 \times 10^{22}$ | $-1.65 \times 10^{22}$ | $-8.02 \times 10^{22}$ | $6.72 \times 10^{22}$ | $1.00 \times 10^{23}$ |
| $b_{235}^{(i)}$ (m$^{-2}$s$^{-2}$) | $9.25 \times 10^{20}$ | $-8.31 \times 10^{22}$ | $-4.23 \times 10^{22}$ | $9.99 \times 10^{22}$ | $1.10 \times 10^{23}$ |
| $b_{244}^{(i)}$ (m$^{-2}$s$^{-2}$) | $6.49 \times 10^{19}$ | $2.24 \times 10^{24}$ | $1.66 \times 10^{20}$ | $-2.46 \times 10^{23}$ | $2.79 \times 10^{20}$ |
| $b_{245}^{(i)}$ (m$^{-2}$s$^{-2}$) | $3.27 \times 10^{22}$ | $1.97 \times 10^{22}$ | $9.85 \times 10^{22}$ | $-6.64 \times 10^{22}$ | $-3.13 \times 10^{23}$ |
| $b_{255}^{(i)}$ (m$^{-2}$s$^{-2}$) | $9.83 \times 10^{19}$ | $3.49 \times 10^{24}$ | $2.80 \times 10^{20}$ | $-1.04 \times 10^{23}$ | $7.20 \times 10^{20}$ |
| $b_{333}^{(i)}$ (m$^{-2}$s$^{-2}$) | $-1.6 \times 10^{22}$ | $1.14 \times 10^{20}$ | $2.73 \times 10^{24}$ | $2.55 \times 10^{20}$ | $-8.13 \times 10^{22}$ |
| $b_{334}^{(i)}$ (m$^{-2}$s$^{-2}$) | $6.78 \times 10^{19}$ | $-2.24 \times 10^{22}$ | $2.94 \times 10^{20}$ | $4.93 \times 10^{24}$ | $3.09 \times 10^{20}$ |
| $b_{335}^{(i)}$ (m$^{-2}$s$^{-2}$) | $-1.29 \times 10^{22}$ | $1.57 \times 10^{20}$ | $-2.56 \times 10^{23}$ | $2.98 \times 10^{20}$ | $7.68 \times 10^{24}$ |
| $b_{344}^{(i)}$ (m$^{-2}$s$^{-2}$) | $-2.41 \times 10^{22}$ | $1.75 \times 10^{20}$ | $4.94 \times 10^{24}$ | $5.21 \times 10^{20}$ | $-7.66 \times 10^{22}$ |
| $b_{345}^{(i)}$ (m$^{-2}$s$^{-2}$) | $-2.98 \times 10^{20}$ | $9.41 \times 10^{22}$ | $3.35 \times 10^{22}$ | $-2.1 \times 10^{23}$ | $-9.32 \times 10^{22}$ |
| $b_{355}^{(i)}$ (m$^{-2}$s$^{-2}$) | $-3.54 \times 10^{22}$ | $2.83 \times 10^{20}$ | $7.72 \times 10^{24}$ | $5.37 \times 10^{20}$ | $-7.11 \times 10^{23}$ |
| $b_{444}^{(i)}$ (m$^{-2}$s$^{-2}$) | $2.19 \times 10^{20}$ | $-7.36 \times 10^{22}$ | $5.59 \times 10^{20}$ | $8.61 \times 10^{24}$ | $1.28 \times 10^{21}$ |
| $b_{445}^{(i)}$ (m$^{-2}$s$^{-2}$) | $-4.03 \times 10^{22}$ | $2.97 \times 10^{20}$ | $-7.65 \times 10^{22}$ | $9.3 \times 10^{20}$ | $1.36 \times 10^{25}$ |
| $b_{455}^{(i)}$ (m$^{-2}$s$^{-2}$) | $2.02 \times 10^{20}$ | $-1.00 \times 10^{23}$ | $5.29 \times 10^{20}$ | $1.36 \times 10^{25}$ | $1.51 \times 10^{21}$ |
| $b_{555}^{(i)}$ (m$^{-2}$s$^{-2}$) | $-6.96 \times 10^{22}$ | $1.23 \times 10^{21}$ | $-2.07 \times 10^{23}$ | $2.53 \times 10^{21}$ | $2.10 \times 10^{25}$ |



\end{document}